\documentclass[10pt,a4paper,twocolumn]{article}
\usepackage{f1000_styles}
\usepackage{float}
\usepackage{wrapfig}
\usepackage{enumitem}

\usepackage[round]{natbib}
\let\cite\citet

\newcommand\aj{{AJ}}%               % Astronomical Journal 
\newcommand\araa{{ARA\&A}}%         % Annual Review of Astron and Astrophys 
\newcommand\apj{{ApJ}}%             % Astrophysical Journal 
\newcommand\apjl{{ApJL}}%           % Astrophysical Journal, Letters 
\newcommand\apjs{{ApJS}}%           % Astrophysical Journal, Supplement 
\newcommand\aap{{A\&A}}%            % Astronomy and Astrophysics 
\newcommand\aapr{{A\&A~Rv}}%        % Astronomy and Astrophysics Reviews 
\newcommand\aaps{{A\&AS}}%          % Astronomy and Astrophysics, Supplement 
\newcommand\mnras{{MNRAS}}%         % Monthly Notices of the RAS 
\newcommand\pasj{{PASJ}}%           % Publications of the ASJ 
\newcommand\ssr{{SSRv}}%            % Space Science Reviews 
\newcommand\nat{{Nature}}%          % Nature 
\newcommand\physrep{{PhR}}%                             % Physics Reports 
\newcommand\pasa{{PASA}}%                               % Publications of the Astron. Soc. of Australia
\graphicspath{ {figures/} }

\begin{document}
\pagestyle{fancy}

\title{Atacama Large Aperture Submillimeter Telescope (AtLAST) science: Gas and dust in nearby galaxies
}
\newcounter{Aff}
\newcounter{NextAff}
\newcounter{NextNextAff}
\newcounter{NextNextNextAff}
\setcounter{Aff}{1}
\setcounter{NextAff}{\theAff}\stepcounter{NextAff}
\author[\theAff,\theNextAff]{%
    Daizhong Liu
    }
    \affil[\theAff]{\normalfont Max-Planck-Institut f\"{u}r extraterrestrische Physik (MPE), Giessenbachstrasse 1,
    %Bayern, 
    D-85748 Garching, Germany}\newcounter{AffMPE}\setcounter{AffMPE}{\theAff}\stepcounter{Aff}
    \affil[\theAff]{\normalfont Purple Mountain Observatory, Chinese Academy of Sciences, 10 Yuanhua Road, Nanjing 210023, China}\stepcounter{Aff}

\setcounter{NextAff}{\theAff}\stepcounter{NextAff}
\author[\theAff,\theNextAff]{%
    Am\'{e}lie Saintonge
    }
    \affil[\theAff]{\normalfont Department of Physics and Astronomy, University College London, Gower Street, London WC1E 6BT, UK}\newcounter{AffUCL}\setcounter{AffUCL}{\theAff}\stepcounter{Aff} 
    \affil[\theAff]{\normalfont Max-Planck-Institut f\"ur Radioastronomie (MPIfR), Auf dem H\"ugel 69, D-53121 Bonn, Germany}\newcounter{AffMPIfR}\setcounter{AffMPIfR}{\theAff}\stepcounter{Aff}

\author[\theAff]{%
    Caroline Bot
    }
    \affil[\theAff]{\normalfont Observatoire Astronomique de Strasbourg, Universit\'{e} de Strasbourg, UMR 7550, 11 rue de l'Universit\'{e}, F-67000 Strasbourg, France}\stepcounter{Aff}

\setcounter{NextAff}{\theAff}\stepcounter{NextAff}
\setcounter{NextNextAff}{\theNextAff}\stepcounter{NextNextAff}
\author[\theAff,\theNextAff,\theNextNextAff]{%
    Francisca Kemper
    }
    \affil[\theAff]{\normalfont Institut de Ci\`{e}ncies de l'Espai (ICE, CSIC), Can Magrans, s/n, E-08193 Cerdanyola del Vall\`{e}s, Barcelona, Spain}
    \stepcounter{Aff}
    \affil[\theAff]{\normalfont ICREA, Pg. Lluís Companys 23, E-08010 Barcelona, Spain}\stepcounter{Aff}
    \affil[\theAff]{\normalfont Institut d'Estudis Espacials de Catalunya (IEEC), E-08860 Castelldefels, Barcelona, Spain}\stepcounter{Aff}

\author[\theAff]{%
    Enrique Lopez-Rodriguez
    }
    \affil[\theAff]{\normalfont Kavli Institute for Particle Astrophysics and Cosmology (KIPAC), Stanford University, Stanford, CA 94305, USA}\stepcounter{Aff}

\author[\theAff]{%
    Matthew W. L. Smith
    }
    \affil[\theAff]{\normalfont School of Physics \& Astronomy, Cardiff University, The Parade, Cardiff CF24 3AA, UK}\newcounter{AffCardiff}\setcounter{AffCardiff}{\theAff}\stepcounter{Aff}

\author[\theAffMPE]{%
    Thomas Stanke
    }

\author[\theAff]{%
    Paola Andreani
    }
    \affil[\theAff]{\normalfont European Southern Observatory, Karl-Schwarzschild-Strasse 2, 85748, Garching, Germany}\newcounter{AffESO}\setcounter{AffESO}{\theAff}\stepcounter{Aff}

\author[\theAff]{%
    Alessandro Boselli
    }
    \affil[\theAff]{\normalfont Aix-Marseille Universit\'{e}, CNRS, CNES, LAM, Marseille, France}\stepcounter{Aff}

\author[\theAff]{%
    Claudia Cicone
    }
    \affil[\theAff]{\normalfont Institute of Theoretical Astrophysics, University of Oslo, PO Box 1029, Blindern, 0315, Oslo, Norway}\newcounter{AffOslo} \setcounter{AffOslo}{\theAff}\stepcounter{Aff}

\author[\theAffCardiff]{%
    Timothy A. Davis
    }

\author[\theAffOslo]{%
    Bendix Hagedorn
    }

\author[\theAffOslo]{%
    Akhil Lasrado
    }

\author[\theAffMPIfR]{%
    Ann Mao
    }

\author[\theAff]{%
    Serena Viti
    }
    \affil[\theAff]{\normalfont Leiden Observatory, Leiden University, P.O. Box 9513, 2300 RA Leiden, The Netherlands}\newcounter{AffLeiden}\setcounter{AffLeiden}{\theAff}\stepcounter{Aff}

\author[\theAff]{%
    Mark Booth
    }
    \affil[\theAff]{\normalfont UK Astronomy Technology Centre, Royal Observatory Edinburgh, Blackford Hill, Edinburgh EH9 3HJ, UK}\newcounter{AffUKRI} \setcounter{AffUKRI}{\theAff}\stepcounter{Aff}

\author[\theAffUKRI]{%
    Pamela Klaassen
    }
    
\author[\theAffESO]{%
    Tony Mroczkowski
    }

\author[\theAff]{%
    Frank Bigiel
    }
    \affil[\theAff]{\normalfont Argelander-Institut f\"ur Astronomie, Universit\"at Bonn, Auf dem H\"ugel 71, D-53121 Bonn, Germany}\stepcounter{Aff}

\setcounter{NextAff}{\theAff}\stepcounter{NextAff}
\author[\theAff,\theNextAff]{%
    M\'elanie Chevance
    }
    \affil[\theAff]{\normalfont Universit\"{a}t Heidelberg, Zentrum f\"{u}r Astronomie, Institut f\"{u}r Theoretische Astrophysik, Albert-Ueberle-Str 2, D-69120 Heidelberg, Germany}\stepcounter{Aff}
    \affil[\theAff]{\normalfont Cosmic Origins Of Life (COOL) Research DAO}\stepcounter{Aff}

\author[\theAff]{%
    Martin A. Cordiner
    }
    \affil[\theAff]{\normalfont Astrochemistry Laboratory, Code 691, NASA Goddard Space Flight Center, Greenbelt, MD 20771, USA}\stepcounter{Aff}

\setcounter{NextAff}{\theAff}\stepcounter{NextAff}
\setcounter{NextNextAff}{\theNextAff}\stepcounter{NextNextAff}
\setcounter{NextNextNextAff}{\theNextNextAff}\stepcounter{NextNextNextAff}
\author[\theAff,\theNextAff,\theNextNextAff,\theNextNextNextAff]{%
    Luca Di Mascolo
    }
    \affil[\theAff]{\normalfont Laboratoire Lagrange, Universit\'e C\^{o}te d'Azur, Observatoire de la C\^{o}te d'Azur, CNRS, Blvd de l'Observatoire, CS 34229, 06304 Nice cedex 4, France}
    \stepcounter{Aff}
    \affil[\theAff]{\normalfont Astronomy Unit, Department of Physics, University of Trieste, via Tiepolo 11, Trieste 34131, Italy}\stepcounter{Aff}
    \affil[\theAff]{\normalfont INAF -- Osservatorio Astronomico di Trieste, via Tiepolo 11, Trieste 34131, Italy}\stepcounter{Aff}
     \affil[\theAff]{\normalfont IFPU -- Institute for Fundamental Physics of the Universe, Via Beirut 2, 34014 Trieste, Italy}\stepcounter{Aff}

\setcounter{NextAff}{\theAff}\stepcounter{NextAff}
\author[\theAff,\theNextAff]{%
    Doug Johnstone
    }
    \affil[\theAff]{\normalfont NRC Herzberg Astronomy and Astrophysics, 5071 West Saanich Rd, Victoria, BC, V9E 2E7, Canada}\stepcounter{Aff} 
    \affil[\theAff]{\normalfont Department of Physics and Astronomy, University of Victoria, Victoria, BC, V8P 5C2, Canada}\stepcounter{Aff}

\setcounter{NextAff}{\theAff}\stepcounter{NextAff}
\author[\theAff,\theNextAff]{%
    Minju M. Lee
    }
    \affil[\theAff]{\normalfont Cosmic Dawn Center, Denmark}\stepcounter{Aff} 
    \affil[\theAff]{\normalfont DTU-Space, Technical University of Denmark, Elektrovej 327, DK2800 Kgs. Lyngby, Denmark}\stepcounter{Aff}

\author[\theAff]{%
    Thomas Maccarone
    }
    \affil[\theAff]{\normalfont Department of Physics \& Astronomy, Texas Tech University, Box 41051, Lubbock TX, 79409-1051, USA}\stepcounter{Aff}

\author[\theAff]{%
    Alexander E. Thelen
    }
    \affil[\theAff]{\normalfont Division of Geological and Planetary Sciences, California Institute of Technology, Pasadena, CA 91125, USA}\stepcounter{Aff}

\author[\theAffESO]{%
    Eelco van Kampen
    }

\author[\theAff]{%
    Sven Wedemeyer
    }
    \affil[\theAff]{\normalfont Rosseland Centre for Solar Physics, Institute of Theoretical Astrophysics, University of Oslo, Postboks 1029 Blindern, N-0315 Oslo, Norway}\stepcounter{Aff}

\maketitle
\thispagestyle{fancy}

\clearpage
\onecolumn

\begin{abstract}

Understanding the physical processes that regulate star formation and galaxy evolution are major areas of activity in modern astrophysics. 
Nearby galaxies offer unique opportunities to inspect interstellar medium (ISM), star formation (SF), radiative, dynamic and magnetic ($\vec{B}$) physics in great detail from sub-galactic (kpc) scales to sub-cloud (sub-pc) scales, from quiescent galaxies to starbursts, and from field galaxies to overdensities. 
In this case study, we discuss the major breakthroughs in this area of research that will be enabled by the Atacama Large Aperture Submillimeter Telescope (AtLAST), a proposed 50-m single-dish submillimeter telescope. The new discovery space of AtLAST comes from its exceptional sensitivity, in particular to extended low surface brightness emission, a very large $2^\circ$ field of view, and correspondingly high mapping efficiency. This paper focuses on four themes which will particularly benefit from AtLAST: 1) the LMC and SMC, 2) extragalactic magnetic fields, 3) the physics and chemistry of the interstellar medium, and 4) star formation and galaxy evolution. With $\sim 1000-2000$~hour surveys each, AtLAST could deliver deep dust continuum maps of the entire LMC and SMC fields at parsec-scale resolution, high-resolution maps of the magnetic field structure, gas density, temperature and composition of the dense and diffuse ISM in $\sim100$ nearby galaxies, as well as the first large-scale blind CO survey in the nearby Universe, delivering molecular gas masses for up to $10^6$ galaxies (3 orders of magnitude more than current samples). Through such observing campaigns, AtLAST will have a profound impact on our understanding of the baryon cycle and star formation across a wide range of environments. 

\end{abstract}

\section*{\color{OREblue}Keywords}

Interstellar medium; Interstellar dust; Magellanic Clouds; Magnetic fields; Astrochemistry; Star formation; Galaxy evolution

\section*{Plain language summary}

The interstellar medium (ISM) -- the gas and dust that permeates galaxies -- governs the evolution of galaxies. Stars form from the coldest regions of the interstellar gas in a complicated process that we still only understand partially. At the same time, the ISM is replenished by material shed by stars and fresh gas accreted from the outside environment.  Building a complete model of star formation requires the study of a wide range of ISM processes on scales ranging from that of an entire galaxy (several kiloparsecs) to that of a small cloud core (below 0.1~parsec). 

With current telescopes, we can however only observe the full details of star formation in our own galaxy.  Nearby galaxies offer a chance to get detailed observations of a more diverse galaxy population, but observing them requires more advanced facilities. The Atacama Large Aperture Sub-millimeter Telescope (AtLAST\footnote{\url{http://atlast-telescope.org/}}) is a 50-m single dish sub-millimeter telescope designed to take on this important challenge. Compared to the current generation of telescopes, AtLAST will be extremely sensitive to faint and extended emission, and able to efficiently map large areas of the sky.

This paper describes four key areas in Nearby Universe research where AtLAST will be particularly impactful.  They are: (1) the study of the dense and diffuse ISM in two of the Milky Way's satellite galaxies, (2) the origin, structure, and physical importance of magnetic fields, (3) the physics and chemistry of the ISM and how it regulates star formation, and (4) the demographics of dust and molecular gas across the galaxy population, from dwarf to giant galaxies.

\clearpage
\pagestyle{fancy}
\twocolumn

%%%%%%%%%%%%%%%%%%%%%%%%%%%%%%%%%%%%%%%%%%%%%%%%%%%%%%%%%%%%%%%%%
%%%%%%%%%%%%%%%%%%%%%%%%%%%%%%%%%%%%%%%%%%%%%%%%%%%%%%%%%%%%%%%%%
%%   INTRODUCTION
%%%%%%%%%%%%%%%%%%%%%%%%%%%%%%%%%%%%%%%%%%%%%%%%%%%%%%%%%%%%%%%%%
%%%%%%%%%%%%%%%%%%%%%%%%%%%%%%%%%%%%%%%%%%%%%%%%%%%%%%%%%%%%%%%%%

\section{Introduction}
\label{sec: introduction}

Of all the important unsolved problems in modern astrophysics, the development of a general and complete theory of star formation, applicable in all environments and cosmic times, would arguably have the greatest impact on our understanding of the universe. The challenge lies in the vast range of scales and processes involved, with gravity, turbulence and magnetic fields being the lead actors \citep[e.g.][]{Mckee2007}. While modern star formation theories can successfully explain the star formation efficiency of molecular clouds in environments similar to the solar neighbourhood, they fail to accurately describe star formation in more extreme environments, such as the high-pressure central regions of galaxies or the low metallicity gas that must have permeated primordial galaxies.

The reason for this predicament is that our understanding of the physics of star formation is currently vastly based on observations of the Milky Way. These observations have the highest physical resolutions possible, but they are limited in the range of environments they can probe (in terms of density, metallicity, UV and X-ray radiation fields, cosmic rays, magnetic fields, etc), not to mention difficulties with distance estimation and confusion along the line of sight.  Nearby galaxies (within $\sim100$~Mpc) are therefore of crucial importance, as they alleviate these issues while still being observable at resolution of tens of parsecs, or even better in the case of Local Group galaxies such as the Small and Large Magellanic Clouds (SMC and LMC, respectively\footnote{The SMC and LMC are the two brightest Milky Way satellite galaxies, and can be seen with the naked eye in the southern sky. They are historically called the Small and the Large Magellanic Clouds, but the usage of the Portuguese explorer's name has a triggering impact on part of the astronomical community and a call has been issued to rename these galaxies as the Small and the Large Milky Cloud. While this matter pends, in this paper we will use their three letters acronyms (SMC and LMC) throughout rather than the full names.}). 

Indeed, significant progress has been made over the past decade thanks to new and upgraded (sub)millimeter facilities such as ALMA and the IRAM 30m/NOEMA starting to enable "Galactic-like" science in extragalactic environments, while also enabling observations of the ISM in large, representative galaxy samples.  From such observations, a picture is emerging where both large-scale environment and local conditions impact the structure and properties of molecular clouds and their star formation outputs: we observe variations from galaxy to galaxy \citep[e.g.][]{Saintonge2022}, with environment within nearby galaxies \citep[e.g.][]{Leroy2021a}, and even between the inner and outer regions of the Milky Way \citep[e.g.][]{Schruba2019}.  In parallel, sensitive dust polarimetry measurements with instruments such as POL2 on the JCMT and HAWC+ on SOFIA are proving beyond doubt the crucial role of magnetic fields in regulating star formation \citep[e.g.][]{Pattle2023,Pattle2019,Pillai2020,Chuss2019}. These studies show that magnetic fields channel gas flows along galactic filaments, thus affecting the structure and history of star formation history on galactic scales.

To achieve another step change in our understanding of ISM physics/chemistry and star formation, we need yet another significant increase in our observing capabilities.  The Atacama Large Aperture Submillimeter Telescope (AtLAST) is a concept for a next generation 50-meter class single-dish submm telescope that would offer such opportunities \citep{Klaassen2020, Mroczkowski2023,Mroczkowski2024}.  AtLAST is designed to offer very high sensitivity and exceptional mapping speeds through both its large aperture, extended 2 degree field of view, and the excellent atmospheric conditions of the Chilean Chajnantor plateau. With its unique combination of spatial coverage, resolution and sensitivity (in particular to extended, low surface brightness emission), AtLAST will enable the study of the ISM in nearby galaxies as has never been possible before.  

In this case study, we explore four prominent nearby galaxies science areas where AtLAST will be transformational: 
\begin{enumerate}[topsep=0pt,itemsep=0ex,partopsep=0ex,parsep=0ex]
\item ISM physics and star formation in the LMC and SMC, with a focus on the important yet poorly studied diffuse gas and dust;
\item the origin and structure of magnetic fields, from tens of kpc down to sub-kpc scales, and their impact on star formation; 
\item the physics and chemistry of the ISM as probed by multiple dust and gas tracers;
\item ISM properties and star formation across the full range of environments of the local Universe through large statistical surveys. 
\end{enumerate}
The background and key science questions are presented in Section \ref{ScienceCase}, highlighting those that AtLAST will be uniquely positioned to address, with a description of possible survey strategies. The technical requirements for AtLAST to reach these science goals are summarised in Section \ref{TechnicalCase}. 
Finally, we summarize the key science goals and technical requirements for AtLAST in Section~\ref{sec: conclusion}.

%%%%%%%%%%%%%%%%%%%%%%%%%%%%%%%%%%%%%%%%%%%%%%%%%%%%%%%%%%%%%%%%%
%%%%%%%%%%%%%%%%%%%%%%%%%%%%%%%%%%%%%%%%%%%%%%%%%%%%%%%%%%%%%%%%%
%%   SCIENCE CASE 
%%%%%%%%%%%%%%%%%%%%%%%%%%%%%%%%%%%%%%%%%%%%%%%%%%%%%%%%%%%%%%%%%
%%%%%%%%%%%%%%%%%%%%%%%%%%%%%%%%%%%%%%%%%%%%%%%%%%%%%%%%%%%%%%%%%

\section{AtLAST nearby galaxies science case highlights}
\label{ScienceCase}

%%%%%%%%%%%%%%%%%%%%%%%%%%%%%%%%%%%%%%%%%%%%%%%%%%%%%%%%%%%%%%%%%
%%%%%%%%%%%%%%%%%%%%%%%%%%%%%%%%%%%%%%%%%%%%%%%%%%%%%%%%%%%%%%%%%
\setlength{\parskip}{6pt}
\subsection{Dust and gas in the LMC and SMC}
\label{sec: MCs}
\subsubsection{Context and open questions}

\begin{figure*}
\centering
\includegraphics[width=0.89\textwidth, trim=0 30mm 0 55mm, clip]{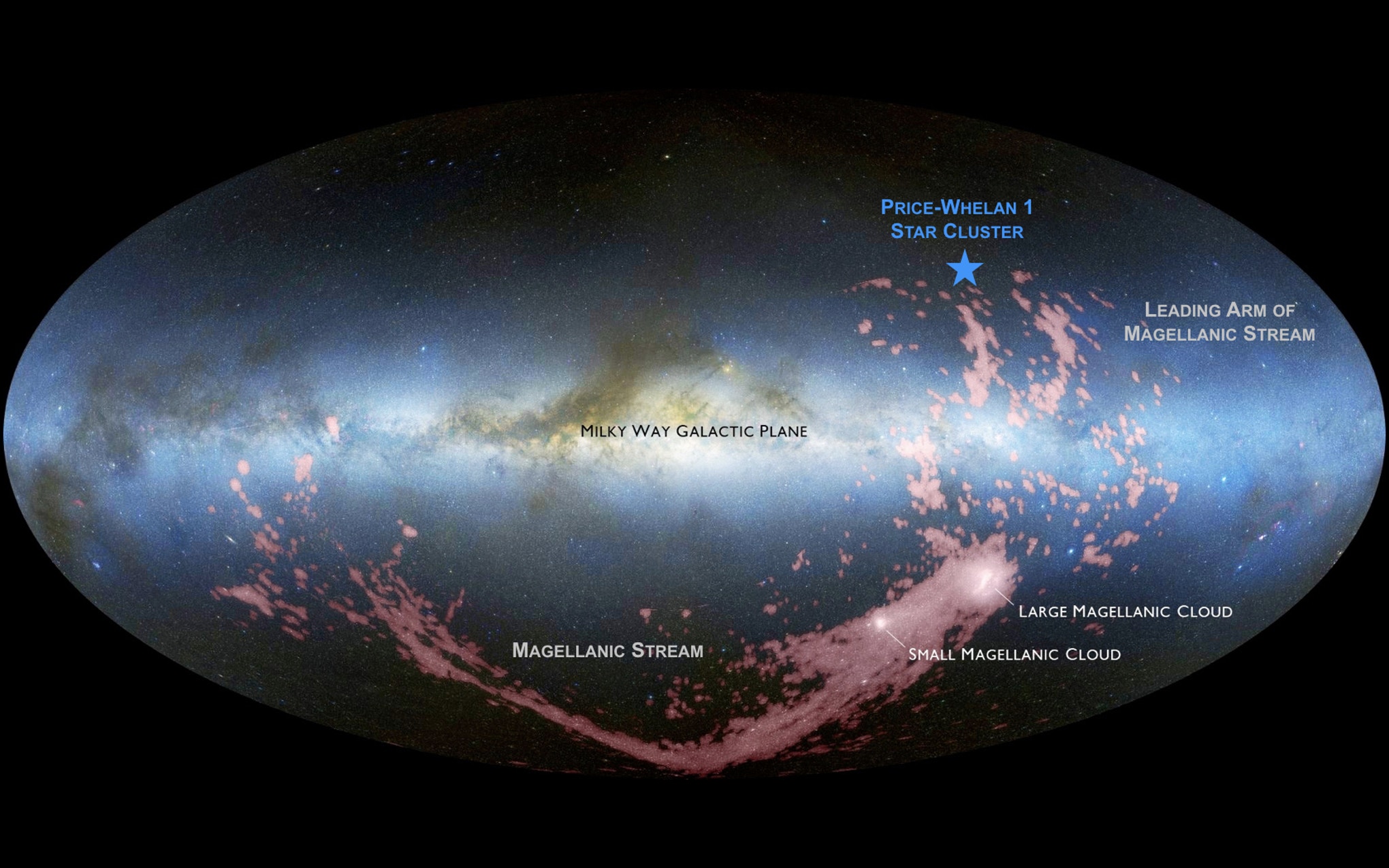}
\includegraphics[width=0.89\textwidth]{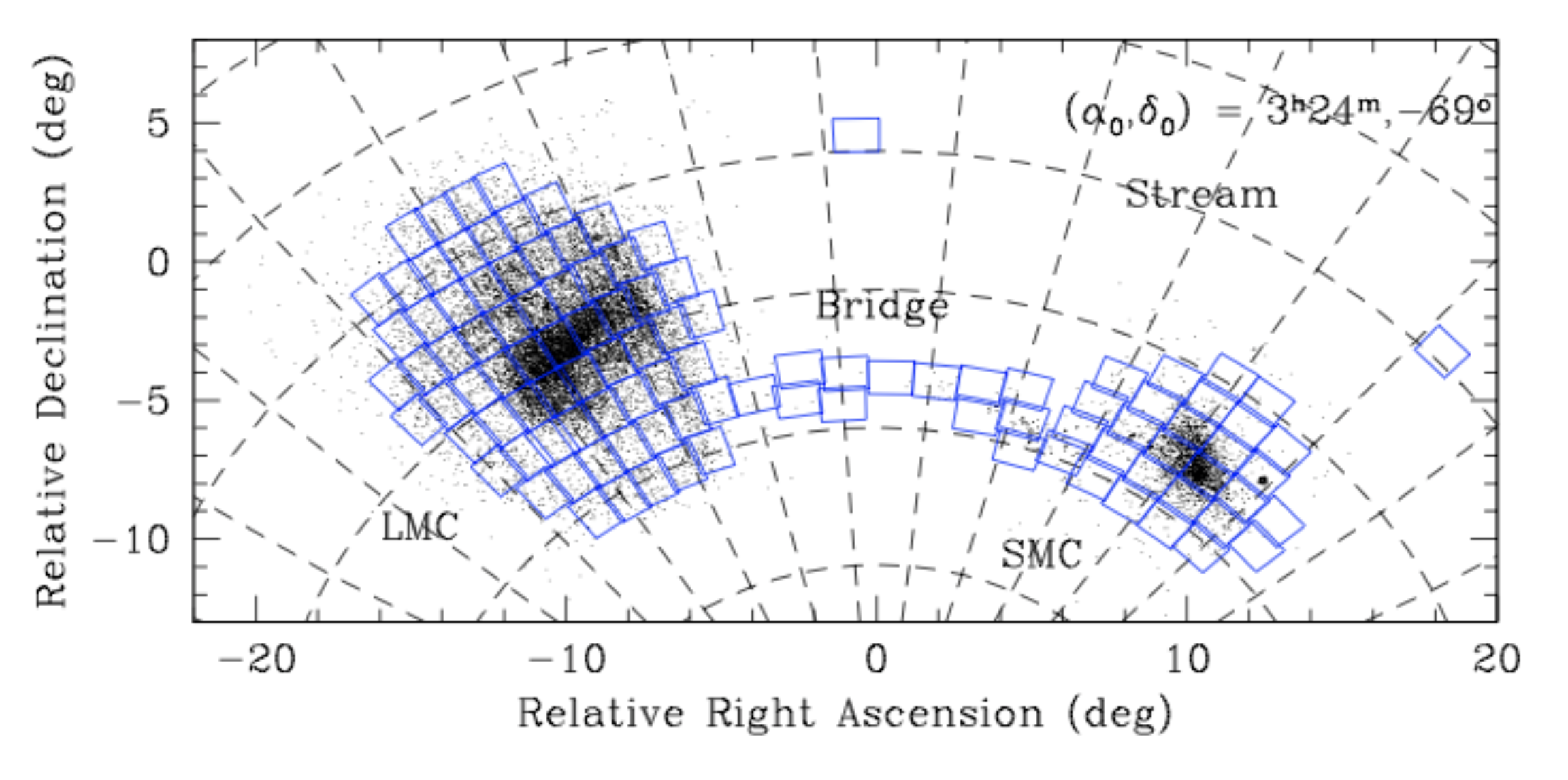}
\caption{%
    Full sky view of the LMC, SMC and the associated Stream system. 
    \textbf{Left:} LMC, SMC, the Stream between the LMC/SMC and the Milky Way, and the leading arm of Stream in H{\sc i} (red), on top of the optical all-sky image in Aitoff projection (West to the left and North up), adapted from \cite{Nidever2010}. 
    (Credit: D. Nidever/ NASA/ Simons Foundation \href{https://earthsky.org/space/milky-way-collision-magellanic-clouds-price-whelan-1/}{URL}.)
    \textbf{Right:} LMC and SMC in sky R.A. and Dec. coordinates (East to the left and North up), with blue squares showing the footprints of the VISTA survey of the LMC/SMC (VMC; \citealt{Cioni2011}).
    \label{fig:full view of magellanic clouds}
}
\end{figure*}

The LMC and SMC are the most massive satellites of our Milky Way Galaxy. Together with Andromeda (M~31), M~33, as well as other smaller galaxies out to $D \sim 1.5\,\mathrm{Mpc}$, they form our Local Group.  The LMC and SMC are strongly bound to the Milky Way and on a collision course, with the LMC expected to merge with our Galaxy in $\sim2.5$~Gyr \citep{Cantun2019}.  The intensity of the gravitational interaction between the SMC, LMC and Milky Way is pulling significant amounts of the interstellar gas away from the galaxies, which can be observed across the sky as large H{\sc i} streams (see Fig.~\ref{fig:full view of magellanic clouds}). 

With gas-phase metallicities of only 20-50\% of the solar abundance, the SMC and LMC are the nearest laboratories to study star formation in low-metallicity environments. This is important, because metallicity is a key parameter regulating the ISM: metals influence the efficiency of gas cooling, the shielding of molecules from the harsh UV radiation field, the production of dust grains, which in turn are the formation sites of molecules. 

At a distance of $\sim$50~kpc \citep{deGrijs2014} and aided by the LMC's low inclination ($i \sim 38^{\circ}$, \citealt{Balbinot2015}) we can spatially resolve the full range of environments present. We can thus unveil physical processes at sub-parsec scales, and quantify the relative contribution of the various ISM components/tracers to the integrated, unresolved ISM observations which are available for distant galaxies. For example, determining locally (in the Milky Way and other very nearby galaxies such as the SMC and LMC) which tracers best measure the amount of high density gas is crucial in estimating the star formation efficiency of any dense gas in large extragalactic samples (see also Sect.\ \ref{sec: line survey}).

The hot and cold ISM of the LMC, both gas and dust, have been extensively observed with radio and (sub)mm facilities such as ASKAP, MeerKAT, ALMA, APEX, and SOFIA (see details in Tables \ref{tab:lmc data sets} and \ref{tab:smc data sets}). In addition, both the LMC and SMC have been observed in the mid- and far-infrared by the {\it Spitzer} SAGE and {\it Herschel} HERITAGE surveys \citep{Meixner2006, Gordon11, Meixner2013}.  Combined, these observations have led to significant new insights, but important questions about the diffuse and dense ISM remain open.

\begin{figure*}
\includegraphics[width=\textwidth]{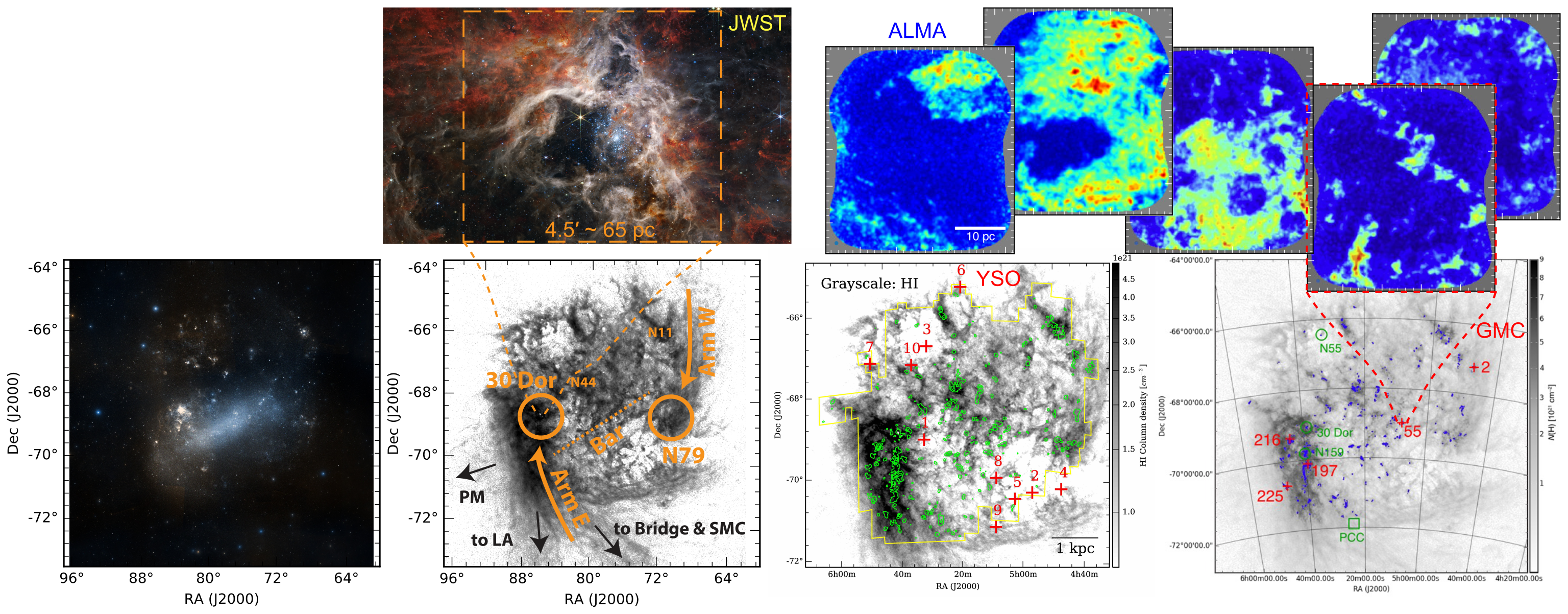}
\caption{%
   \textbf{Bottom four panels:} Stellar light, H{\sc i}, and CO emission in the LMC. Panels from left to right show the optical RGB image from \href{https://sky.esa.int/esasky/}{the ESA Sky}; H{\sc i} emission in gray with labels of regions from \cite{Ochsendorf2017}; CO(1--0) emission (green) from the NANTEN telescope (at $2.6' \sim 40$~pc resolution; \citealt{Fukui2008}) on top of H{\sc i} column density gray image (at $\sim 14' \sim 200$~pc resolution; \citealt{Kim2003}) adapted from \cite{Harada2019}; and the ALMA small mosaic observations at a few locations highlighted in green and red, adapted from \cite{Sawada2018}. 
   \textbf{Top-left panel} shows the JWST NIRCam RGB image of the Tarantula Nebula star-forming region at ultra-high resolution ($0.06'' \sim 3000$~AU), adapted from \href{https://www.nasa.gov/universe/a-cosmic-tarantula-caught-by-nasas-webb/}{NASA website}. 
   \textbf{Top-right five panels} show the ALMA observations of the $^{12}$CO(1--0) emission in five GMCs in the LMC (field of view $2.5' \times 2.5'$ each) at sub-parsec scales (at $3'' \sim 0.7$pc resolution) from \cite{Sawada2018}. 
   \label{fig:GMCs}
}
\end{figure*}

\paragraph{How much molecular gas is there?}
These existing observing campaigns yield important constraints on the properties of the ISM in the LMC and SMC (see Fig.~\ref{fig:GMCs}).  For example, \cite{Fukui2008} show that most of the molecular gas in the LMC is contained in over a hundred individual giant molecular clouds (GMCs; \citealt{Hughes2010}), with a mean molecular mass surface density $\sim 50 \, \mathrm{M_{\odot}\,pc^{-2}}$, making those clouds only half as dense as those in the inner Milky Way. On the other hand, observations of the [CII] line emission in the LMC with SOFIA/GREAT suggest that $\gtrsim 75\%$ of the cold gas may be CO-dark \citep{Chevance2020}. Accurately quantifying the molecular gas contents is of foremost importance, as a significant open question in the field is whether the ability of an individual GMC to form stars (i.e. its star formation efficiency) is determined by the intrinsic properties of the cloud, or dependent on interstellar environments \citep[e.g.][]{Krumholz2005,Girichidis2020,Chevance2020Review}. The first step to answer this fundamental question is therefore to characterise the molecular gas content (both CO-dark and CO-bright) of GMCs of various properties in a broad range of environments characterised by different levels of large-scale turbulence, pressure, external shear, and magnetic field strengths \citep[e.g.][]{Andre2017,Pattle2023}.

\paragraph{How much dust is there?}

Using common methods and calibrations, it can be demonstrated that the interstellar dust masses calculated for the SMC are too large to be explained by the dust production from evolved stars \citep{Schneider2014, Srinivasan2016}, assuming that dust destruction and star formation are non-negligible sinks of interstellar dust in the ISM. Similarly, excessively large dust masses are also observed in high redshift (and therefore low metallicity) galaxies, with these observations particularly hard to explain at $z\gtrsim 6$, when the Universe is too young for Asymptotic Giant Branch stars to have evolved to the dust forming stage \citep[e.g.,][]{Dwek2011,Rowlands2014,Witstok2023}. 

However, recent work on dust in the LMC, SMC and other nearby resolved galaxies suggests that this ``dust budget crisis'' might be attributed to our lack of understanding of cosmic dust in low density, low metallicity environments and wrongly applying assumptions based on observations of dust in high metallicity environments. Indeed, submm and mm emission excesses have been observed in nearby galaxies, especially at low metallicities and/or in diffuse environments such as the ISM of the LMC \citep{Galliano2011}, and even more prominently in the SMC \citep{Bot2010, Planck2011, Gordon2014}. This excess can be interpreted as an indication of a different dust composition, grain properties or emissivity at longer wavelengths in low metallicity environments \citep{Galliano2018}, or the emission from magnetic nanoparticles \citep{Draine2013}. Furthermore, \citet{Clark2021} have shown that in such large galaxies, \textit{Herschel} observations suffered from spatial filtering on large scales, compounding the problem. After correcting for this missing large scale, diffuse emission by adding data from the Plank satellites, part of the tensions between observations and models however appear to be resolved \citep{Clark2023}.  This highlights the vital importance of getting  sub-mm and mm observations at a resolution that resolves individual environments and that also account for diffuse dust emission, if we are to characterise the properties of interstellar dust in low metallicity environments.

\subsubsection{The role of AtLAST in the study of the LMC and SMC}
\label{sec: MCs with AtLAST}

Interferometers like ALMA and current single dish submillimeter telescopes like APEX are very inefficient at mapping large-scale, low-density interstellar medium at multiple bands, either due to filtering of flux on large scales, poor sensitivity, poor spatial resolution, or low mapping efficiency. The characterization of the diffuse molecular and atomic ISM, where GMCs are born and remain embedded, is therefore impossible without a next-generation (sub)millimeter single dish telescope such as AtLAST.  We give an overview here of how AtLAST will tackle the core questions raised above regarding the ISM in the LMC/SMC, together with a description of the corresponding technical requirements in Section \ref{sec: MCs technical discussion}.

\paragraph{What are the properties of interstellar dust in the diffuse and dense ISM?}

So far, studies of the distribution and properties of dust in the LMC/SMC have relied heavily on the full maps observed with {\it Herschel} at wavelengths of $\sim 70$--500~$\mu$m (see summary in Tables \ref{tab:lmc data sets} and \ref{tab:smc data sets}). To expand on these results, higher spatial resolution measurements are required to trace spatial variations in dust properties, being sensitive to large scale emission is required to properly trace the very diffuse dust component, and longer wavelength data are required to increase the precision to which we can measure the dust emissivity index (and as a consequence dust temperature) and the dust surface density. The advantages of such an approach are demonstrated for instance for M~31, which was targeted by the HASHTAG programme with SCUBA-2 on the JCMT at 450 and 850~$\mu$m \citep{Smith2021}. Unfortunately, it is currently not possible to make large scale continuum maps ($7^{\circ} \times 7^{\circ}$ for the LMC and $3^{\circ} \times 5^{\circ}$ for the SMC) at submm or mm wavelengths. The exception is the survey performed with the South Pole Telescope, which was expanded to include the LMC and SMC at 1.4, 2.1 and 3 mm  \citep{Crawford2016}. However, this survey has a spatial resolution of only 1.0$'$ to 1.7$'$ ($\sim 400$--800~pc), and contains contributions from free-free and synchrotron radiation complicating the measurement of thermal dust emission at these wavelengths. 

With AtLAST, we will characterise the shape of the dust spectral energy distribution from the far-infrared to the millimeter in several bands, from dense regions to diffuse environments. This will make it possible to separate the cold dust component from free-free emission, but also from background emission from distant galaxies or CMB fluctuations. Studying this emission in the two different metallicity regimes offered by the LMC and SMC will enable us to understand how dust varies with environment and hence to model it precisely and ultimately get better dust properties (including the most precise dust temperature, emissivity index $\beta$, and dust masses). The significantly (up to $\sim 15 \times$) higher spatial resolution compared to \textit{Herschel} will make it possible to trace substructures, and characterise the multi-temperature dusty medium.

\paragraph{How can we trace dark molecular gas, and what is its importance?}

As metallicity decreases, CO is restricted to the highest density parts of the molecular clouds, leaving a substantial molecular layer of H$_2$ without CO that has been dubbed "CO-dark" \citep[e.g.][]{Wolfire10}. This CO dark component is often conceptualised as the outer layer of giant molecular complexes, but in reality the geometry of the ISM is much more complicated. Given the filamentary structure of the ISM, only small CO-emitting clumps survive the photoionisation, making them very difficult to detect without sensitive, high resolution and large field-of-view observations \citep[as observed for example with ALMA in the low metallicity dwarf galaxy WLM,][]{Rubio2015}.  Observations of the [CII] line emission in the LMC with SOFIA/GREAT suggest that $\gtrsim 75\%$ of the cold gas may be CO-dark \citep{Chevance2020}.  To reach our goal of understanding how star formation proceeds in low density, low metallicity environments, it is vital that we accurately resolve the molecular gas reservoirs of galaxies such as the SMC and LMC, and therefore we must look beyond CO for molecular ISM tracers. 

The neutral carbon's [CI](1--0) and (2--1) fine structure lines are popular tracers of CO-dark gas, proposed to be very promising for tracing the CO dissociation layer of PDRs and the low-metallicity gas without much CO \citep{Glover2016, Ramambason2024}. By observing both [CI] lines, it is furthermore possible to measure the temperature, density and optical depth of the gas. Significant mapping of the [CI] in the LMC and SMC is currently impossible: ALMA's field of view is far too small, and a 12m single dish like APEX does not offer enough sensitivity and resolution. A survey of the LMC and SMCs with AtLAST in Bands 8 and 9\footnote{Throughout this review, we adopt the ALMA definition of observing frequency bands, see \url{https://www.eso.org/public/teles-instr/alma/receiver-bands/}} would therefore allow us to map not just the distribution of the diffuse molecular gas, but its physical conditions as well, at unprecedented sensitivity and resolution.  This would be a game-changer in our understanding of the ISM and of star formation in diffuse, low metallicity extragalactic environments. For example, accurate mapping of the full molecular gas reservoir (CO {\it and} CO-dark) will allow us to answer the very important question of the extent to which star formation can occur in diffuse ISM environments. 

\paragraph{How does metallicity drive the gas-to-dust ratio?}

Given the low metallicity ISM of especially the SMC, key physical and chemical processes that regulate cooling, heating, dust and molecule formation are different than in the Milky Way.  While we could hope that such processes scale linearly with the amount of metals available, studies of nearby galaxies have shown that reality is more complex. For example, the gas-to-dust ratio is observed to evolve linearly as expected only at high metallicities.  Below a critical metallicity threshold, the amount of dust per gas unit breaks down from the expected relationship \citep[e.g.][]{remyruyer14, romanduval2017,Galliano2018}. This critical metallicity of $\sim 0.2 Z_{\odot}$ is usually interpreted as the threshold above which grain growth becomes efficient \citep{Zhukovska2016}. An alternative interpretation is that the break corresponds to a regime change where diffuse gas becomes more important \citep{Clark2023}. Straddling this metallicity transition region, the LMC/SMC offer us a unique opportunity to accurately measure both dust and gas abundances to significantly improve our understanding of how these two components of the ISM interrelate.  AtLAST will be essential in achieving these goals, as they can only be achieved through a complete census of the metals (e.g., with the optical integral-field Local Volume Mapper survey; \citealt{Kollmeier2019}), dust and gas (with AtLAST).

\subsubsection{AtLAST surveys of the LMC and SMC}
\label{sec: MCs technical discussion}

At declinations of $\sim -70$~degrees, the LMC and SMC are only observable from the Southern hemisphere (see Fig.~\ref{fig:full view of magellanic clouds}). They have already been mapped by a wide range of radio and (sub)mm telescopes, as summarised in Tables \ref{tab:lmc data sets} and \ref{tab:smc data sets}, but suffer from lack of resolution, confusion, and/or lack of sensitivity to large-scale flux.  AtLAST will be transformational, by allowing full mapping of the LMC and SMC over nearly a hundred square degrees in continuum emission and moderately bright lines, scanning through the electromagnetic spectrum from Band~3 ($\sim$ 100 GHz) to Band~10 ($\sim$ 850 GHz) to determine accurate dust and gas properties. The best spatial resolution can be achieved in the highest-frequency Band~10, about 1.5--2~arcsec (0.4--0.6~pc).

\paragraph{Continuum observations}

With AtLAST we will be able to obtain multi-band submillimeter continuum maps with sufficient sensitivity and area coverage to trace emission from bright star forming regions to the faint, diffuse ISM. Coverage in multiple bands will be crucial: it is needed in the first place to derive the spectral shape of the dust emission to constrain the dust composition and grain sizes, and to discriminate between dust and free-free emission in regions where ionized gas is important. Due to the sensitivity and large field of view of AtLAST, "contamination" by CMB fluctuations will be an issue, but with multi-band observations we will be able to disentangle them, given the well characterised spatial and spectral characteristics of the CMB. Similarly, we can expect to be affected by confusion with more distant sub-millimeter/millimeter galaxies, resolved or unresolved (Cosmic Infrared Background), which will create correlated noise that will have to be estimated, but can be separated from dust emission using its spatial scale properties and wavelet phase harmonics \citep[][]{Auclair2024}.

\begin{figure*}
    \centering
    \includegraphics[width=0.47\textwidth]{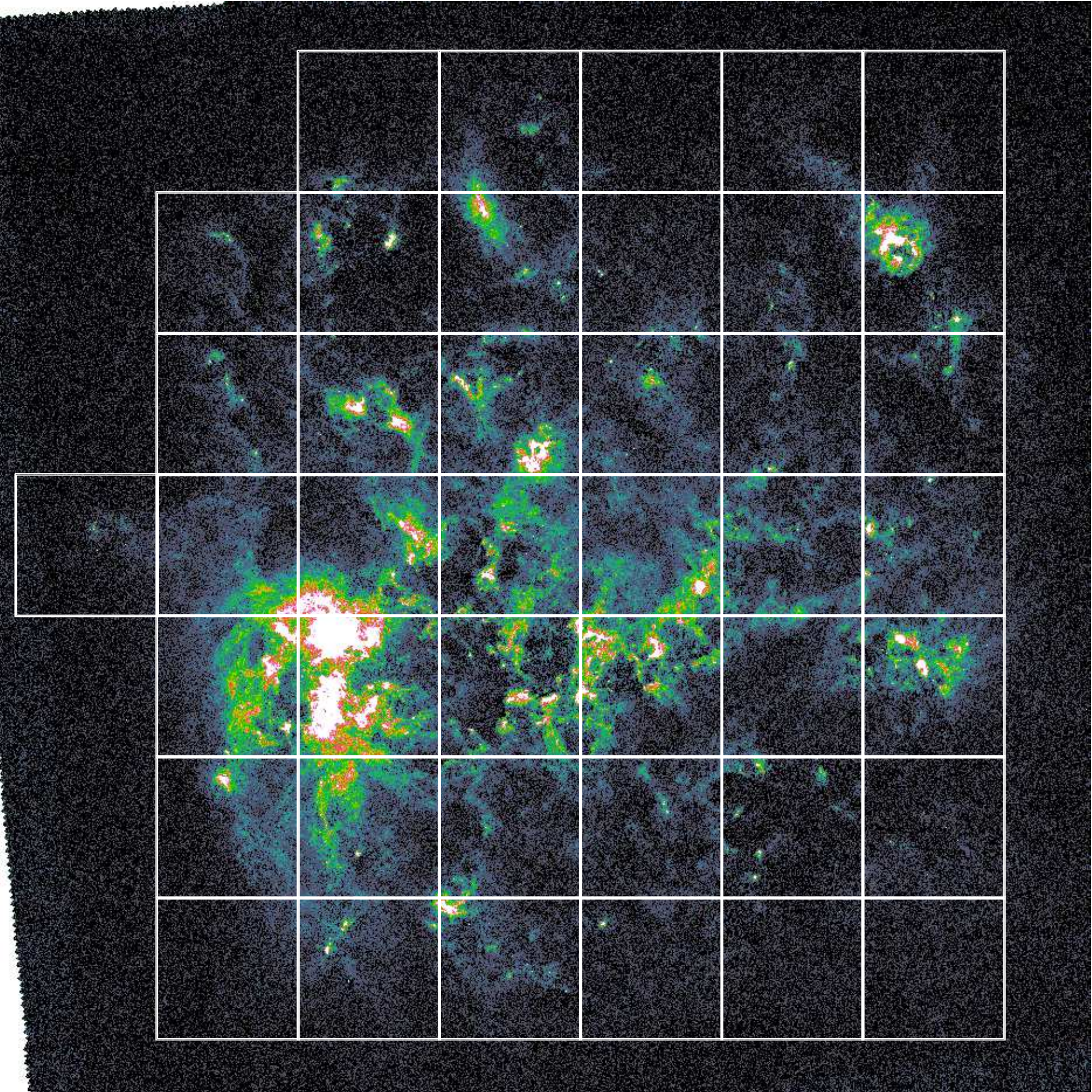}
    \includegraphics[width=0.47\textwidth]{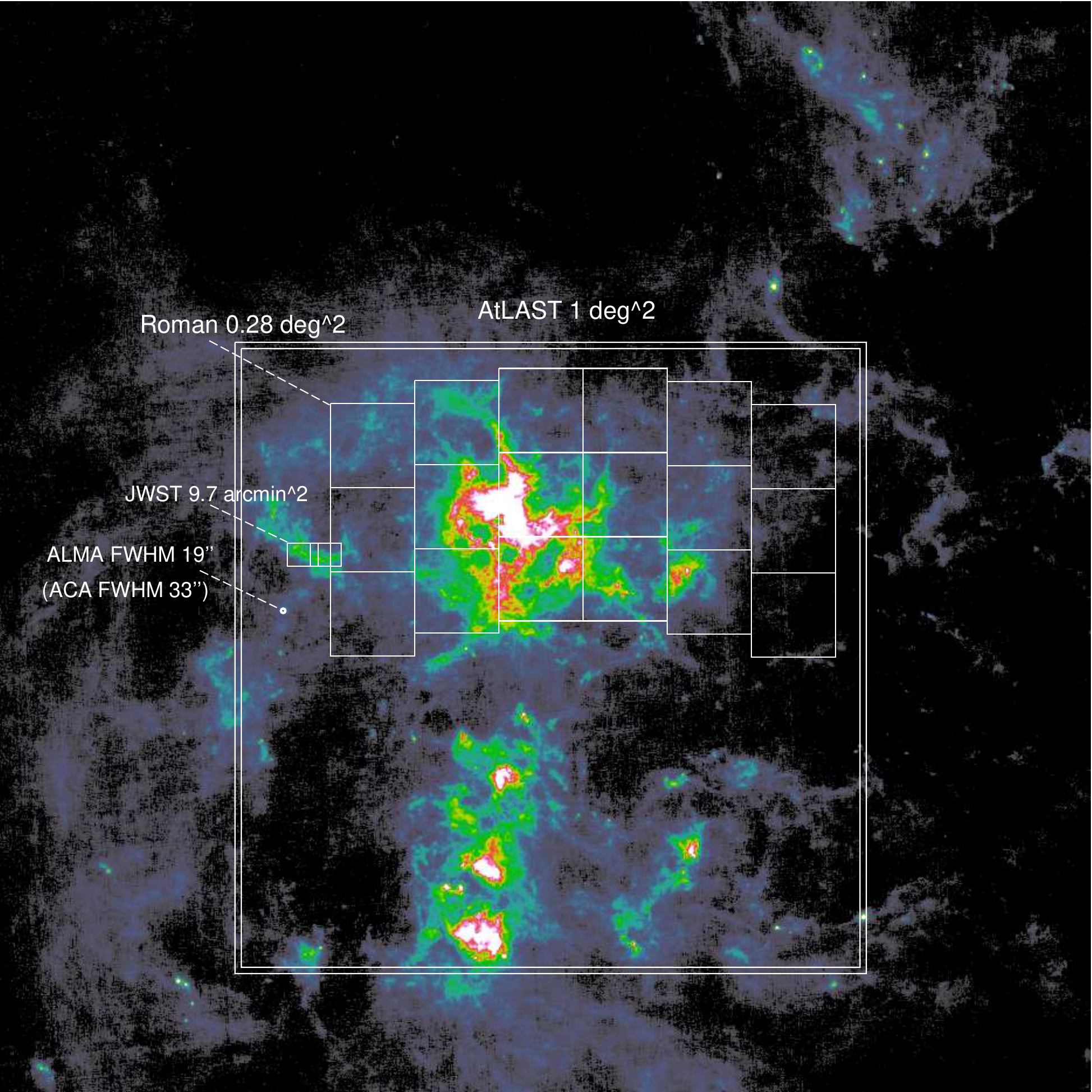}
    \caption{Dust emission at 100~$\mu$m in the LMC from the \textit{Herschel} space observatory \citep{Meixner2013} The left panel shows the full $\sim 7^{\circ} \times 7^{\circ}$ area of LMC. Grids are $1^{\circ} \times 1^{\circ}$ fields of view of AtLAST continuum camera. The right panel shows the zoom-in $30'$ view around the 30 Doradus H{\textsc{ii}} region. The fields of view of AtLAST, ALMA (12-m array and 7-m array), \textit{JWST} NIRCam and \textit{Roman} Space Telescope WFI are shown for comparison. Only the future-generation AtLAST and \textit{Roman} are capable to map the full near to far-infrared and sub-millimeter dust emission in the LMC.}
    \label{fig:lmc-herschel}
\end{figure*}

\begin{figure*}
    \centering
    \includegraphics[width=0.47\textwidth]{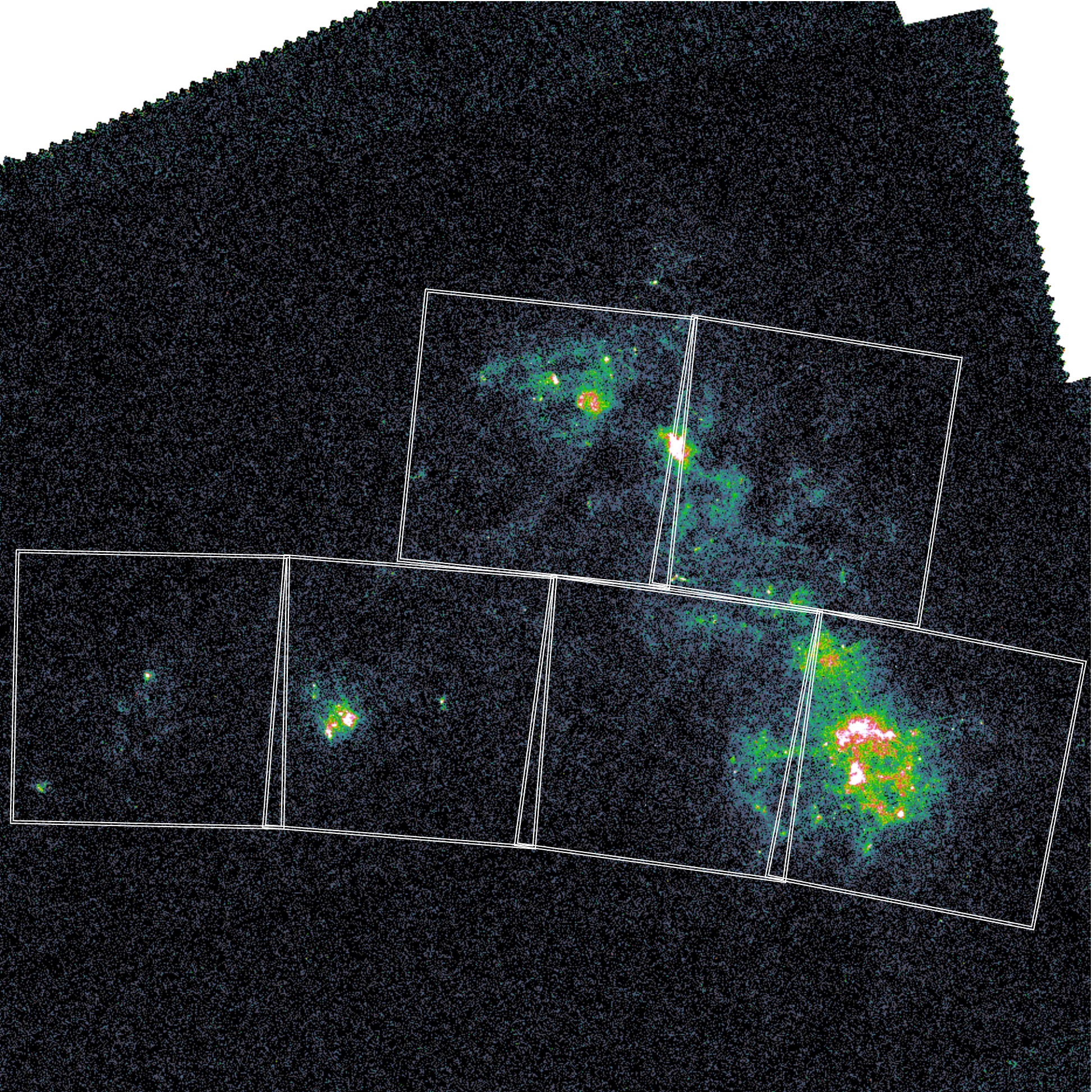}
    \includegraphics[width=0.47\textwidth]{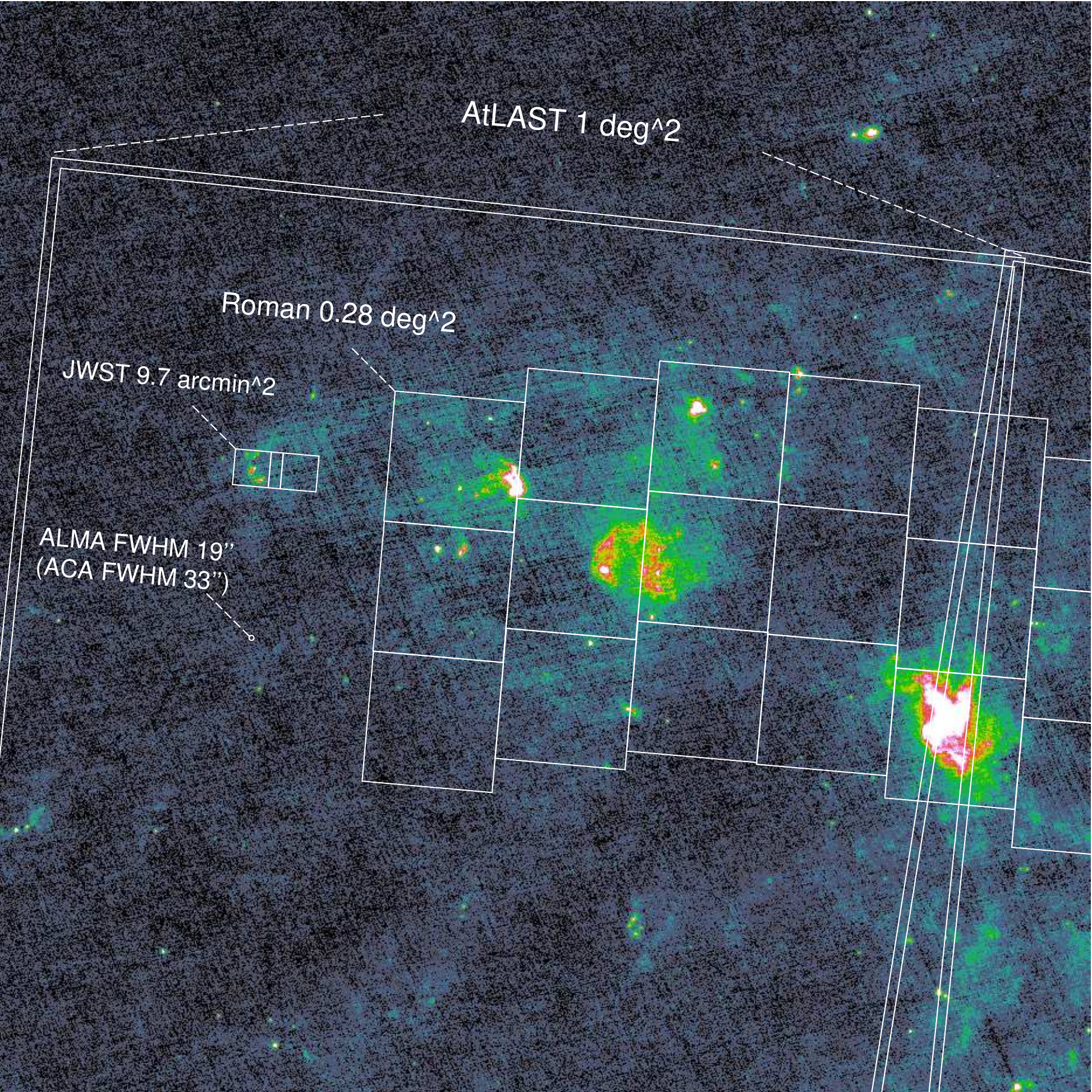}
    \caption{%
    Similar to Fig.~\ref{fig:lmc-herschel} but for the SMC. 
    }
    \label{fig:smc-herschel}
\end{figure*}

\begin{table*}
    \centering
    \begin{tabular}{p{3cm}|p{10cm}}
    \hline
    \hline
        ISM phase & LMC  \\
    \hline
        Atomic gas (HI) 
            & ATCA+Parkes, beam$\sim1'$ \citep{Kim2003,StaveleySmith2003} \\
            & GASKAP: an HI survey with the ASKAP telescope \citep{Pingel2022} \\
            & LGL-BS: a VLA extra-large L-Band survey (\href{https://www.lglbs.org/home}{URL}) \\
            & MagiKAT: a MeerKAT large survey of HI and OH \citep{vanLoon2010} \\
        Molecular gas (CO)
            & CO(1--0) at beam$\sim 8.8'$ over $6^{\circ} \times 6^{\circ}$ with Columbia 1.2-m Millimeter-Wave Telescope \citep{Cohen1988} \\
            & Various CO(1--0) observations in selected regions with the SEST 15-m (Israel et al. 1986, 2003; Johansson et al. 1994, 1998; Caldwell \& Kutner 1996; Kutner et al. 1997) \\
            & NANTEN complete survey of CO(1--0) at beam$\sim 2.6'$ with NANTEN 4-m, covering $\sim 30$~deg$^2$ \citep{Fukui1999,Fukui2008} \\
            & MAGMA survey of CO(1--0) at beam$\sim 33''$ with Mopra 22-m, covering 3.6~deg$^2$ \citep{Wong2011,Pineda2009,Hughes2010} \\
            & APEX CO(2--1) at beam$\sim 30''$ in 3--13~arcmin$^2$ areas \citep{Okada2019} \\
            & APEX CO(3--2) at beam$\sim 20''$ ($\sim 5$~pc) over 13.8~deg$^2$ \citep{Grishunin2023} \\
        PDR tracers ([CI]) 
            & AST/RO in N159/N160 H~{\sc ii} regions at beam$\sim 3.8'$ \citep{Bolatto2000c} \\
            & Herschel/HIFI at $\sim 44''$ for [CI](1--0) and $\sim 26.6''$ for [CI](2--1) along 36 lines of sight \citep{Pineda2017} \\
        Ionized gas \& PDR ([CII]) 
            & Herschel/HIFI at $\sim 12''$ along 36 lines of sight \citep{Pineda2017} \\
            & SOFIA/GREAT at $\sim 16''$ in 3--13~arcmin$^2$ areas \citep{Okada2019} \\
            & SOFIA observations \citep{Madden2023} \\
        Ionized gas \& H~{\sc ii} regions (H$\alpha$)
            &  H$\alpha$ maps \citep[MCELS][]{Smith1998} \\
        Dust (IR/submm)  
            & Spitzer SAGE \citep{Meixner2006}: all 7 IRAC and MIPS bands \\
            & Herschel HERITAGE \citep{Meixner2013}: beam$\sim 7'' \text{--} 36''$, area$\sim 7^{\circ} \times 7^{\circ}$ \\
            & QUEST maps from \citet{Clark2021} that includes the diffuse dust from a combination with Planck \\
            & high resolution dust reddening at 25$''$ resolution from Chen et al 2020 \\
            & South Pole Telescope (with Planck) \citep{Crawford2016} \\
            & SOFIA/HAWC+ dust continuum and polarization at 53-214 $\mu$m with beam =5-18$''$ \citep{Tram2021,Tram2023} \\
    \hline
    \hline
    \end{tabular}
    \caption{Observations of ISM tracers in the LMC.
    }
    \label{tab:lmc data sets}
\end{table*}

\begin{table*}
    \centering
    \begin{tabular}{p{3cm}|p{10cm}}
    \hline
    \hline
        ISM phase & SMC  \\
    \hline
        Atomic gas (HI) 
            & HI survey with ATCA+Parkes at beam$\sim1.6'$ (30~pc)\citet{Stanimirovic1999} \\
            & GASKAP: an HI survey with the ASKAP telescope \citep{Pingel2022,Dempsey2022} \\
            & LGL-BS: a VLA extra-large L-Band survey (\href{https://www.lglbs.org/home}{URL}) \\
            & MagiKAT: a MeerKAT large survey of HI and OH \citep{vanLoon2010} \\
            & HI outflows with ASKAP at $\sim30''$ \citep{McClureGriffiths2018,Dempsey2020} \\
        Molecular gas (CO)
            & CO(1--0) at $\sim8.8'$ ($\sim 160$~pc) with Columbia 1.2-m Millimeter-Wave Telescope \citep{Rubio1991} \\
            & CO(1--0) and $^{13}$CO(1--0) at $\sim45''$ with SEST \citep{Israel1993} \\
            & CO(1--0) at $\sim 43''$ with SEST \citep{Rubio1993} \\
            & CO(2--1) at $\sim 22''$ and CO(1--0) at $\sim43''$ with SEST \citep{Rubio1996} \\
            & CO(1--0) at $\sim 42''$ with the Mopra 22-m \citep{Muller2010} \\
            & CO(1--0), $^{13}$CO(1--0) and C$^{18}$O(1--0) at $\sim 33''$ with Mopra 22-m along 18 lines of sight, plus CO(3--2) and $^{13}$CO(3--2) at $\sim 17.5''$ with APEX 12-m \citep{Pineda2017} \\
            & CO(2--1) at $\sim 27''$ ($\sim 9$~pc) with APEX \citep{DiTeodoro2019b} \\
            & CO(2--1) at $\sim 27''$ ($\sim 9$~pc) in a 0.47~kpc$^2$ region with APEX \citep{Saldano2023} \\
            & CO(3--2) at $\sim 20''$ ($\sim 6$~pc) in $\sim 2.5$~deg$^2$ region with APEX \citep{Saldano2024} \\
        Radio 
            & MeerKAT 1.3~GHz \citep{Cotton2024} \\
        PDR tracers ([CI]) 
            & ISO satellite at $3.8'$ \citep{Bolatto2000b} \\
            & Herschel/HIFI at $44''$ for [CI](1--0) and $26.6''$ for [CI](2--1) along 18 lines of sight \citep{Pineda2017} \\
        Ionized gas \& PDR ([CII]) 
            & Herschel/HIFI at $12''$ along 18 lines of sight \citep{Pineda2017} \\
        Ionized gas \& H~{\sc ii} regions (H$\alpha$)
            & H$\alpha$ maps \citep[MCELS][]{Winkler2015} \\
        Dust (IR/submm) 
            & Spitzer SAGE \citep{Meixner2006}: all 7 IRAC and MIPS bands \\
            & Herschel HERITAGE \citep{Meixner2013}: beam$\sim 7'' \text{--} 36''$, area$\sim 7^{\circ} \times 7^{\circ}$ \\
            & South Pole Telescope (with Planck) \citep{Crawford2016} \\
            & QUEST maps from \citet{Clark2021} \\
            & AzTEC 1.1~mm continuum over 4.5 deg$^2$ with ASTE 10-m \citep{Takekoshi2017} \\
    \hline
    \hline
    \end{tabular}
    \caption{Observations of ISM tracers in the SMC.
    }
    \label{tab:smc data sets}
\end{table*}

Figures~\ref{fig:lmc-herschel} and \ref{fig:smc-herschel} show the footprint of a proposed mapping survey of the LMC and SMC with AtLAST.  Assuming a field of view of 1~deg$^2$ for the first-generation AtLAST continuum camera, we will need $\sim 40$ pointings to cover the LMC ($\sim 7^{\circ} \times 7^{\circ}$) and $\sim 8$ pointings to cover the SMC ($\sim 4^{\circ} \times 2^{\circ}$). For each pointing, the integration time depends on the bandwidth and depth. Here we adopt a high-spatial-resolution multi-wavelength SMC dust simulation by C.\ Bot et al. (in prep.) to estimate the expected observing time. The simulation uses the dust parameters from \citet{Clark2023} to extrapolate the observed dust spectral energy from the \textit{Herschel} HERITAGE survey \citep{Meixner2013} combined with Planck on large scales \citep{Clark2021} to AtLAST wavelengths and sampled at the AtLAST angular resolutions (Table~\ref{tab:atlast-angular-resolutions}). CMB fluctuations contributing on large scales were also added using the SMICA map from Planck data \citep{Planck2020}.
 From these maps, we find that the detection limit can be set by the faintest dust structures in the SMC bar region, which has flux densities of the order of 0.01~MJy/sr at 100~GHz (Band 3), to 0.25~MJy/sr at 400~GHz (Band 8) and 0.5~MJy/sr at 680~GHz (Band 9). 
Given the angular resolution of AtLAST at these bands, we can derive the RMS to achieve at least $3\sigma$ per angular resolution unit: which is 1.5~$\mu$Jy/beam, 22~$\mu$Jy/beam and 19~$\mu$Jy/beam at 100~GHz, 400~GHz and 680~GHz, respectively. 

Using the \href{https://www.atlast.uio.no/sensitivity-calculator/}{AtLAST sensitivity calculator}, with a bandwidth of 32~GHz, elevation 45~deg, and H$_2$O percentiles of 70, 40 and 10 for the aforementioned bands, we obtain an on-source integration time of 27.0~h, 5.2~h, 28.5~h, respectively, for one pointing. 
The Band~8 observation (calculated at 400~GHz) has the best observing efficiency in terms of the shortest integration time to achieve the detection limit. This is set by both the dust SED shape and the angular resolution at that frequency. 
Other bands are more time-consuming, but are still crucial to constrain the shape of dust SED at each position. 
To map the whole LMC and SMC, we need to multiply the above times by the number of pointings. Thus the total amount of on-source time will be about $\sim 2900$~h. Spreading the observing time into multiple years, and in each year to obtain a full mapping of LMC and SMC, will be a wise option to ease the scheduling pressure and also to allow for transient studies (see companion case study by Orlowski-Scherer et al., in prep.). 

This AtLAST LMC/SMC survey will eventually provide multi-band (e.g., 3 bands, from Band 3 $\sim 100$~GHz to Band 9 $\sim 700$~GHz) deep dust continuum mapping, with angular resolutions $\gtrsim 10 \times$ better than \textit{Herschel} and $\gtrsim 4 \times$ better than APEX, with unprecedented sensitivity to diffuse emission that ALMA will completely miss, and with important frequency and spatial coverages not covered by \textit{Herschel}, ALMA, APEX, LMT, etc. 

\paragraph{Line observations}
The molecular gas in the LMC has (in an area-covering way) so far only been observed in CO main isotopologue lines, e.g., in the CO(1-0) line by \citet{Fukui1999,Fukui2008} and \citet{Kawamura2009} with the NANTEN 4-m telescope at 40~pc (2.6$'$), and further at 11~pc (30$''$) by \citet{Wong2011} with the Mopra 22-m telescope for selected small regions (3.6~deg$^2$ in total). 
An on-going large survey with the APEX 12-m telescope targets the $^{12}$CO(3--2) and $^{13}$CO(3--2) lines at 5~pc resolution, aiming to cover 17.4~deg$^2$ or 13.3~kpc$^2$ \citep{Weiss2023,Grishunin2023}. Deeper and/or higher spatial resolution observations of CO and other tracers remain limited to individual pointings or small specific regions (e.g., \citealt{Wong2017,Wong2022,Sawada2018}). 
The lack of contiguous, sensitive, high-resolution multi-line coverage of at least significant portions of the LMC and SMC, in order to sample the full range of environments they contain, is still a very important obstacle preventing us to understand the physical and chemical structure of our neighbour low-metallicity galaxies. 

We here outline prospects (and limitations) for possible line mapping surveys of the LMC and SMC with AtLAST. First, we describe a CO(3--2) survey that would correspond to the APEX $^{12}$CO(3--2)/$^{13}$CO(3--2) survey of \cite{Weiss2023} and \cite{Grishunin2023}, but 4$\times$ higher spatial resolution and an order of magnitude higher sensitivity. Such a survey would reveal CO-bright molecular gas in a much more complete way (including intrinsically faint structures as well as small features suffering from beam dilution if observed with poorer spatial resolution). It also serves as an illustration as to which extent extensive mapping of correspondingly fainter lines from other tracers will be possible. Second, we describe a survey in the neutral carbon $[$CI$]$ lines at 492 and 809~GHz, which will be key in understanding the contribution of "CO-dark" molecular gas and atomic gas at the transition between the diffuse ionized ISM and molecular clouds.
For the $[$CI$]$ surveys we set the target sensitivity to be able to achieve 4-5~$\sigma$ detections of a 0.1~K line brightness, as the $[$CI$]$ lines  detected by \cite{Pineda2017} and \cite{Bolatto2000c} are at levels of a few times 0.1~K (with line-widths of typically a few km/s).

\begin{table*}
    {
    \centering
    \begin{tabular}{p{3cm}|p{2cm}|p{2.8cm}|p{2.6cm}|p{1.8cm}|p{2.3cm}}
    \hline
    \hline
    Survey & frequency & resolution & sensitivity & pwv H$_2$O & time/ \\
           & (GHz)  & spatial/velocity & 1~$\sigma$ & percentile & square degree\\
    \hline
    Wei\ss{}2023 & 345.8, 330.6 & 19$^{\prime\prime}$/0.5~km/s & 200~mK & & 52~h $^{(a)}$ \\
    AtLAST $^{13}$CO & 330.6 & 4.6$^{\prime\prime}$/0.5~km/s & 20~mK/38~mJy & 40 & $\sim$ 40~h $^{(b)}$\\
    AtLAST $[$CI$]$~$^3$P$_1$--$^3$P$_0$ & 492.16 & 3$^{\prime\prime}$/1.0~km/s & 22~mK/40~mJy & 20 & $\sim$260~h $^{(b)}$ \\
     &  &  &  & 10 & $\sim$150~h $^{(b)}$ \\
     &  &  &  & 40 & $\sim$1000~h $^{(b)}$ \\
    AtLAST $[$CI$]$~$^3$P$_2$--$^3$P$_1$ & 809.34 & 1.8$^{\prime\prime}$/1.0~km/s & 22~mK/40~mJy & 10 & $\sim$1500~h $^{(b)}$ \\
    \hline
    \end{tabular}
    \caption{Telescope time needed to cover one square degree for the APEX CO(3-2) survey of \cite{Weiss2023} (Wei\ss{}2023) using the 7-pixel LASMA receiver array, a corresponding deeper AtLAST survey at similar frequencies (AtLAST $^{13}$CO), and higher frequency AtLAST $[$CI$]$ surveys (assuming 1000 pixel heterodyne arrays on a 50~m diameter AtLAST). AtLAST telescope times were estimated using the
    \href{https://www.atlast.uio.no/sensitivity-calculator/}{AtLAST sensitivity calculator}, assuming an elevation of 40$^\circ$ (accounting for the low declination of the LMC/SMC).}
    \label{tab:LMC_SMC_linesurveys}
    }
    \footnotesize{$^{(a)}$ including observational overheads}
    \footnotesize{$^{(b)}$ without any observational overheads}
\end{table*}

Table \ref{tab:LMC_SMC_linesurveys} summarizes our example LMC/SMC line surveys in terms of the time that would be required to cover one square degree. For the "APEX-like" CO(3-2) survey, AtLAST (if equipped with a large multi-pixel heterodyne array) will provide mapping speeds comparable with contemporary facilities, but with an order of magnitude improvement in sensitivity and a factor of several in spatial resolution. Full cloud scale maps in a number of lines significantly fainter than the main CO lines will be possible, on scales ranging from the full galaxies down to about one parsec, i.e., the scales of clumps within clouds that are thought to form clusters or groups of stars.

At higher frequencies, aiming for lines such as the $[$CI$]$ lines or higher-$J$ CO lines, the situation is more challenging. It will still be possible to cover significant area at sufficient sensitivity in the brightest lines (particularly if observations can be spread over several observing periods) at moderately high frequencies (e.g., $[$CI$]$~$^3$P$_1$--$^3$P$_0$ in ALMA Band~8). At even higher frequencies (ALMA Bands~9 and 10) observations will be restricted to partial coverage. In terms of $[$CI$]$~$^3$P$_2$--$^3$P$_1$ observations (required to constrain the excitation conditions of the gas), we will likely need to adopt the observing strategy of targeting only regions where the $^3$P$_1$--$^3$P$_0$ line has been detected.

Table \ref{tab:LMC_SMC_linesurveys} also shows the time estimate for the $[$CI$]$~$^3$P$_1$--$^3$P$_0$ 492~GHz line for three different values of the H$_2$O pwv percentile. While at the 10 and 20 percentile (corresponding to about 0.5 and 0.8~mm pwv, respectively, according to the documentation of the ETC) significant (multi-square degree) coverage is still feasible, this clearly becomes prohibitive at the 40 percentile level (corresponding to about 1.5~mm pwv). Going to a high and dry site, as AtLAST will, is absolutely mandatory if any significant sensitive coverage at these frequencies is to be achieved.

%%%%%%%%%%%%%%%%%%%%%%%%%%%%%%%%%%%%%%%%%%%%%%%%%%%%%%%%%%%%%%%%%
%%%%%%%%%%%%%%%%%%%%%%%%%%%%%%%%%%%%%%%%%%%%%%%%%%%%%%%%%%%%%%%%%
\vspace{0.5in}
\subsection{Extragalactic magnetic fields}
\subsubsection{Context and open questions}

Cosmic magnetic fields are one of the most elusive forces in our Universe, even though their importance has become increasingly clear in all areas of astrophysics. Magnetic fields are one of the main regulators of local star formation efficiency \citep{Krumholz2019} and play a crucial role in the ISM dynamics in galaxies \citep{Beck2015}. They affect the transport of ionized gas, dust, and metals outward to the circumgalactic medium (CGM) via galactic outflows \citep{Thompson2006,LopezRodriguez2021a,LopezRodriguez2023a,LopezRodriguez2023c}, and they are central to the dynamics of cosmic rays \citep{RodriguezMontero2023}. B-fields are a crucial component in theoretical galaxy formation \citep[e.g.,][]{Moss2000,Hennebelle2014,MartinAlvarez2020,Hopkins2023}, with the new generation of numerical magnetohydrodynamical (MHD) simulations (e.g. \cite{Wissing2022}) posing magnetic fields and cosmic rays as alleviating forces for challenges such as matching the halo mass-stellar mass or mass-metallicity relations.

Magnetic fields in galaxies are thought to have originated as a weak seed during the early stages of the Universe \citep{Widrow2002,Subramanian2019}. These magnetic fields have to be amplified during galaxy formation to explain the observed $\sim\mu$G strengths in present-day galaxies \citep{Beck2019}. Turbulence-driven dynamos are expected to be responsible for the generation of most of the magnetic energy required to reach these magnetic field strengths \citep{Subramanian2019}. Dynamos convert kinetic energy into magnetic energy. Mergers within the cosmic web trigger considerable star formation activity through starbursts. During these starburst phases, the magnetic fields of galaxies can be further enhanced via turbulent dynamos driven by stellar feedback such as supernova (SN) explosions. This generates tangled B-fields at and below the turbulent scale of approximately $<50-100$ pc for SN feedback \citep{Brandenburg2005,Haverkorn2008}. All of this leads to a theoretical scaling of $B_{\rm{tot}} \propto SFR^{1/3}$ in models of a galaxy dominated by SN-driven turbulence \citep{Schober2013}. Thus, an amplification of the B-fields from the early universe to present-day galaxies is expected to be driven by star formation activity. 

Most of our knowledge about magnetic fields in galaxies has been established using radio polarimetric observations in the 3-20 cm wavelength range \citep{Beck2013}. These observations trace the magnetic fields via synchrotron emission and Faraday rotation measurement arising from high-energy particles passing through a magneto-ionic ISM in the warm and diffuse phase of the ISM \citep{MartinAlvarez2023}. This method is widely used to measure the magnetic field strengths and directions in our Milky Way as well as nearby galaxies. Magnetic  field strengths are estimated assuming equipartition between magnetic energy and cosmic ray electrons. Measuring the Rotation Measure (RM) requires at least three frequency bands with large frequency separations to accurately estimate the magnetic field direction along the LOS. These remarkable efforts have shown that all spiral galaxies have kpc-scale ordered magnetic fields with an average strength of 5$\pm$2 $\mu$G \citep{Beck2019}. The total magnetic field strength, measured by synchrotron total intensity and assuming equipartition between the total B-field and total cosmic-ray electron density, is estimated to be $\sim$17$\pm$14 $\mu$G \citep{Fletcher2010,Beck2019}. However, these measurements have been performed at low resolution ($>$15$''$), which corresponds to a spatial scale of $\sim300$ pc at 20 Mpc, and tracing the warm ISM at vertical heights $>200$ pc from the galaxy's midplane, missing the information about the small-scale magnetic field closely related to the star formation activity. The  SKAO will provide sub-arcsecond resolutions, thus enabling complementary high-resolution L-/C-band radio polarization studies, although the small-scale Faraday-rotation and Faraday-depolarization due to foreground clouds of diffuse thermal gas and magnetic fields can be very challenging to account for, for instance, the so-called ``canals'' in the maps of polarized intensity \citep[e.g.,][]{Haverkorn2004a,Haverkorn2004b}.

Far-infrared/submillimeter polarimetry, on the other hand, observes the polarized thermal emission from magnetically aligned dust grains associated with the cold and dense phase of the ISM in galaxies. Elongated dust grains irradiated by starlight spin along the axis of their greatest moment of of inertia (i.e., the minor axis of the dust grain) producing a magnetic moment that aligns with the local magnetic field in the ISM. Thus, polarized thermal emission produce a measurable polarization with their position angle perpendicular to the local B-field.  This technique has successfully traced the magnetic fields in more than a dozen nearby (<20 Mpc) galaxies (i.e., spirals, starbursts, dwarfs, mergers, active galactic nuclei) in the 53-250 $\mu$m using SOFIA/HAWC+ at resolutions of 4.8-18$''$ (90 pc - 1 kpc) (see most of the latest results from the \href{http://galmagfields.com/}{SOFIA SALSA large survey}; \citealt{LopezRodriguez2022b,Borlaff2023}), as well as in starburst galaxies, M82 at $850$ $\mu$m with the JCMT at a resolution of 15$''$ (277 pc) \citep{Pattle2021} and NGC~253 at $860$ $\mu$m with ALMA at a resolution of $\sim0.3''$ (5 pc) \citep{LopezRodriguez2023c}. FIR polarimetric observations of galaxies trace the magnetic fields as a density-weighted average within the beam size sensitive to the densest gas in the cold neutral phase of the ISM \citep{MartinAlvarez2023}. Specifically, SOFIA/HAWC+ observations trace FIR magnetic fields associated with dense, $\log_{10} (N_{\rm{H}} [\rm{cm}^{-2}])=19-23$, and cold, $T_{\rm{d}} = 19-48$ K, dusty, turbulent star-forming regions (those traced at FIR) and these are less ordered than those in the warmer, less dense interstellar medium (those traced at radio) \citep{LopezRodriguez2022b,Borlaff2023}. 

FIR polarimetric observations show that all spiral galaxies have kpc-scale magnetic fields spatially correlated with the molecular gas and with large angular dispersion (i.e., turbulent magnetic field) arising from the star-forming regions in the spiral arms \citep{Borlaff2021,LopezRodriguez2021b,LopezRodriguez2023a}. SOFIA/HAWC+ $154$ $\mu$m observations of the Antennae, a merger system of two spiral galaxies, showed a circumgalactic medium permeated with a $9$~kpc scale ordered magnetic field connecting both pair members  \citep{LopezRodriguez2023a}. This large-scale ordered magnetic field is not detected at radio wavelengths due to Faraday depolarization and the star formation activity in this region. In addition, FIR polarimetric observations are sensitive to the magnetic fields in the cold phase of the galactic outflows in starburst galaxies \citep{LopezRodriguez2021a,LopezRodriguez2023c}. Note that radio polarimetric observations suffer from the short lifetime of CR and Faraday rotation in the galactic outflows challenging the characterization of magnetic fields in starburst galaxies \citep{Adebahr2017}. For M82, the magnetic field strength in the cold phase of the galactic outflow was estimated to be $305\pm15~\mu$G at a resolution of $90$ pc (4.8$"$) at $53$ $\mu$m. This magnetic field was estimated to be `open' into the CGM enriching it with dust metals and astrophysical magnetic fields amplified by supernovae explosions. The magnetic field strength was estimated using a modified version of the Davis-Chandrasekhar-Fermi (DCF) method \citep{Davis1951,CF1953}. The original DCF was developed to measure the magnetic field strength for a steady state and uncompressible medium using the measurements of the gas velocity dispersion and the dispersion of the polarization angles. The modified DCF method includes large-scale flows and shearing to account of the several components in the galactic disk and galactic outflow \citep{LopezRodriguez2021a,Guerra2023}. However, the low resolution observations achieved by SOFIA and JCMT impede the quantification of the magnetic energy in galaxies because the turbulent coherence length of the magnetic field is not resolved. Therefore, the energy budget in the ISM of host galaxies responsible for the gas dynamics and star formation history is still unknown, and the theoretical
scaling of $B_{\rm{tot}} \propto SFR^{1/3}$ is still unconfirmed.

The FIR/sub-mm polarized spectrum provides unique information about the properties of dust and physical conditions of the star-forming regions. In the Milky Way, the polarization spectrum of star forming regions has been observed to be wavelength dependent in the  FIR/sub-mm regime \citep[e.g.,][]{Hildebrand1999,Vaillancourt2008,Michail2021}. This dependency is thought to arise from temperature gradients where only the warm silicate dust grains with high emissivity index embedded in the star-forming regions are aligned, while a colder component mainly following a black body emission is not aligned. Although dust models \citep[e.g.,][]{Guillet2018,HD2023} have been able to explain the polarized spectrum in the diffuse ISM of the Milky Way, these models do not reproduce the wavelength dependence of the polarized spectrum in regions of strong radiation fields in the Milky Way or nearby starburst galaxies. The diffuse ISM is characterized by a roughly flat polarized spectrum with a homogeneous mix of dust components. However, the strong radiation fields in star-forming regions can change the dust composition, dust grain sizes, and dust temperatures. In nearby starbursts, the unresolved gradient of temperatures and tangled magnetic field along the LOS makes the interpretation of the polarized spectrum very complex \citep{LopezRodriguez2023c}. However, the shape of the polarized spectrum has been shown useful to distinguish between disk- and starburst-dominated galaxies for unresolved polarimetric observations, and to characterize the dust composition on the cold phase of the galactic outflows. Multi-wavelength polarimetric observations is required to characterize the dust composition in the ISM of galaxies.

Thus, despite all these efforts so-far, we still do not fully understand how the measured B-fields in galaxies originated, are amplified, or are affected by star formation activity in galaxies. This is mainly due to the fact that radio wavelengths trace the warm ISM in a volume-filling medium and suffer from Faraday rotation and short lifetimes of cosmic rays in dense regions. The B-fields in the cold and dense ISM, where most of the mass and star formation reside, have been overlooked. 

Even when SKA becomes available to map a large number of nearby galaxies' B-fields at (sub)arcsecond scales, we will still lack the crucial high-resolution far-infrared/submillimeter view of the magnetic fields. The ALMA interferometer can help us to some extent in those efforts, however, it is very inefficient in surveying a large number of galaxies, and more importantly, missing substantial diffuse ISM where magnetic fields play a crucial role. A complete view of magnetic fields in nearby galaxies, as a benchmark to understand distant Universe as well as ISM and MHD physics, is something AtLAST can uniquely deliver.  

\begin{figure*}
    \centering
    \includegraphics[width=0.95\linewidth]{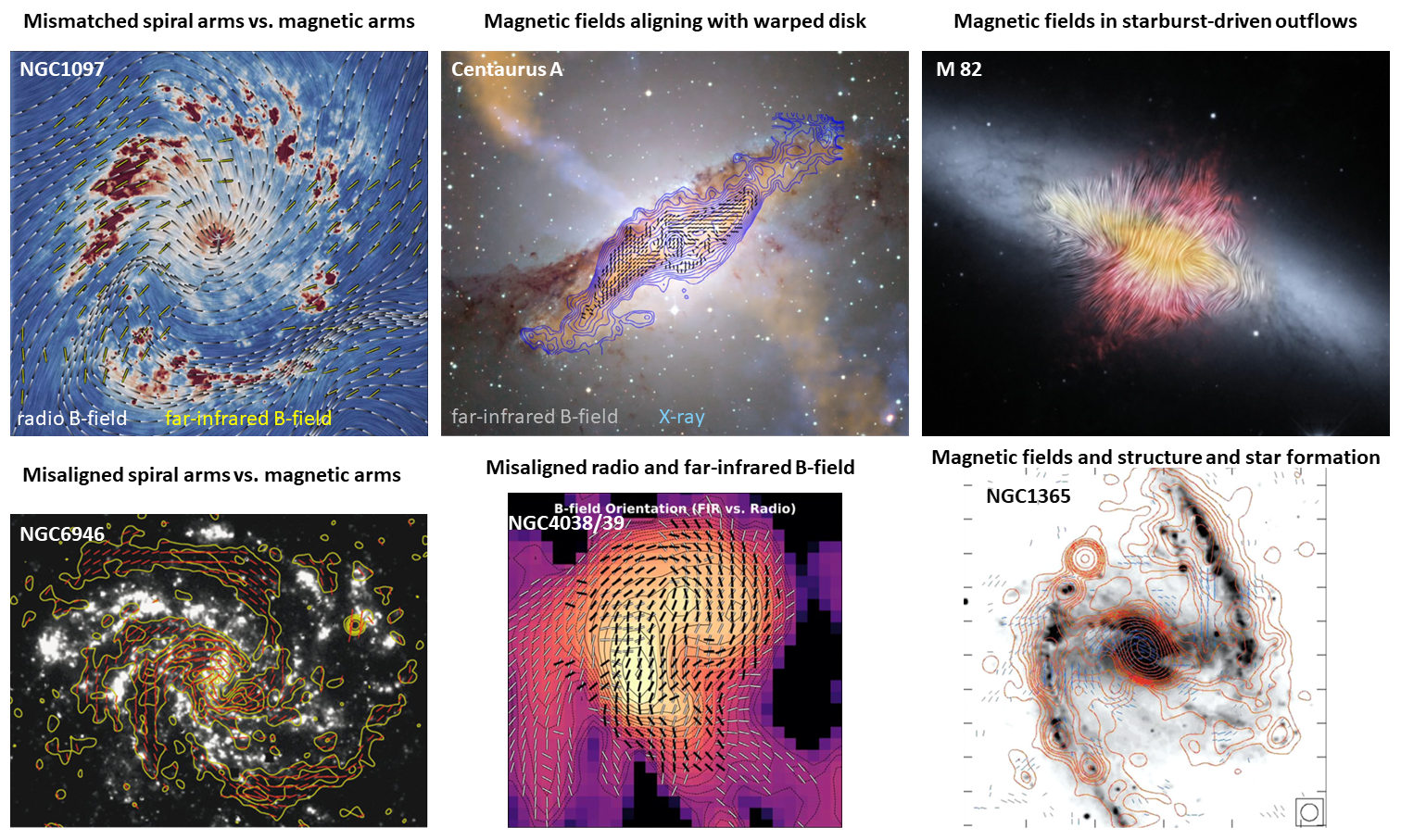}
    \caption{%
    Magnetic fields in nearby galaxies and puzzling questions at a glance. Figures in the six panels from left to right and top to bottom are adapted from the following references: \citet{LopezRodriguez2021c}, \citet{LopezRodriguez2021b}, \citet{LopezRodriguez2023c}, \citet{Beck2013}, \citet{LopezRodriguez2023a} and \citet{Beck2013}, respectively. 
    \label{fig:magnetic-fields}
    }
\end{figure*}

\subsubsection{The role of AtLAST in the study of extragalactic magnetic fields}

The concept for the AtLAST 50-m single dish submillimeter telescope \citep{Mroczkowski2024} has transformational capabilities for characterizing and understanding magnetic fields in our local Universe. First is its unprecedented survey speed for a large number ($\sim$~hundreds) of galaxies thanks to the large field of view. Nearby ($<20$ Mpc ) galaxies are typically tens of arcmin  in diameter (except for the LMC/SMC and M31), which can be rapidly covered by AtLAST using a single pointing ($\sim1$ deg$^{2}$). Note that ALMA requires multiple pointings to cover a single galaxy making it very inefficient and even for the brightest galaxies, unfeasible. Second is its superb sensitivity to map the faint, diffuse dust emission in nearby galaxies and their circumgalactic media. This extended and diffuse emission is filtered out by interferometers. Third is its capability of observing the high-frequency multi-bands achieving 1.5--3$''$ resolution to resolve nearby (<20 Mpc) galaxies at giant molecular cloud scales (20--300~pc). These resolutions perfectly match existing and future large surveys of nearby galaxies at cloud scales (e.g., the PHANGS large program; \citealt{Leroy2021b}; the EDGE-CALIFA survey; \citealt{Bolatto2017}, etc.), which allows a comprehensive study of the multi-phase ISM and correlates the magnetic field with the gas flows in galaxies.

Taking full advantage of AtLAST in observing nearby galaxies at a few arcsecond resolution, the following key science questions will be eventually addressed (to name a few): 

\begin{itemize}[topsep=0pt,itemsep=0ex,partopsep=0ex,parsep=1ex]
    \item {{How do magnetic fields affect star formation at cloud scales?}}     
    \item {{Does the star formation efficiency correlate with the magnetic field strength?}}     
    \item {{Is the magnetic field the support of supersonic turbulence in the ISM?}}
    \item {{What determines the diversity and evolution of magnetic field strengths among galaxies?}}
\end{itemize}

\subsubsection{AtLAST surveys to map magnetic fields in nearby galaxies}

In addition to their immediate scientific  objectives, polarization studies simultaneously will provide very deep and multi-wavelength dust continuum observations. For example, to achieve an uncertainty of $0.3$\% in the degree of polarization a signal-to-noise ratio of $\sim500$ in the total intensity is required. Therefore through polarimetric observations,  deep dust continuum maps will be provided to the community to perform more general studies of dust emission and dust grain properties.  

To enable significant statistics, a survey of dust polarization in over a hundred nearby galaxies with AtLAST 50m is needed. Note that each galaxy will be resolved at scales $20-300$ pc providing thousands of statistically independent polarization measurements that will be used to study trends between the magnetic field structure and strength with the star formation activity and gas kinematics in the disk of galaxies.

We assume a conservative polarization of 1\% by means of magnetically aligned dust grains in resolved polarimetric observations of nearby galaxies, based on the thermal dust polarization in spiral galaxies at 53--220 $\mu$m \citep{LopezRodriguez2022a} and that the polarization is roughly constant in the disk of galaxies in the 50--800 $\mu$m wavelength range  \citep{LopezRodriguez2023c}. Star-forming regions show a wavelength dependence in the polarized spectrum in the FIR/sub-mm regime with a minimum polarization of $~1$\% in the $100-200 \mu$m range and increasing to $\sim4$\% at shorter and longer wavelengths. The shape of the polarized spectrum is still under active study as it provides unique information about dust properties and composition. Thus, we estimate the required observing time to map $\sim 100$ nearby galaxies at the fiducial frequencies of 230~GHz (Band 6), 345~GHz (Band 7), 460~GHz (Band 8), 690~GHz (Band 9) and 806~GHz (Band 10) with AtLAST 50-m. 

To estimate the sensitivity requirements, we use as a reference the JCMT SCUBA-2 HASHTAG/DOWSING large programs which reach a rms noise level of $\sim 2.0$~mJy/beam (0.41~MJy/sr) with a $13.5''$ beam at 850\,$\mu$m \citep{Smith2021}.  Given the large SCUBA-2 beam,it is expected that the dust emission filling factor is $\ll 1$, resulting into a flux density that would be much higher when measured with the smaller AtLAST beam. 
We use the PHANGS-ALMA CO(2--1) $1''$ imaging data \citep{Leroy2021a} of a typical spiral galaxy NGC~4321 (M~100) to estimate the beam filling factor. We smooth the CO moment-0 image to the $3.3''$ AtLAST beam and to the $13.5''$ SCUBA-2 beam, respectively, then examine the emission at the outer disk and find a beam filling factor of 0.2. Furthermore, considering a greybody dust spectral energy distribution of $S/S_{850\mu\mathrm{m}} \propto \nu/\nu_{850\mu\mathrm{m}}^{3.8}$, we can estimate an SCUBA-2-equivalent RMS~$5.7$~MJy/sr ($\sim 4.5$~mJy/beam) with the AtLAST beam at 460~GHz. In order to detect the 1\% polarized dust emission, we need to go 100$\times$ deeper than the SCUBA-2 surveys with AtLAST. 
Thus an RMS of $\sim 16$~$\mu$Jy/beam is needed. 

We use the \href{https://www.atlast.uio.no/sensitivity-calculator/}{AtLAST sensitivity calculator} to estimate the required on-source integration time. Assuming a source elevation of 45~deg, observing frequency 460~GHz, bandwidth 32~GHz, single polarization sensitivity, and H$_2$O profile percentile of 20 (30), we need 28.6~h (47.3~h) on-source integration time per galaxy.  Similarly, at 345~GHz and 660~GHz (RMS~$\sim 10$~$\mu$Jy/beam and $\sim 31$~$\mu$Jy/beam), with 50-th and 20-th H$_2$O percentile, we need 18~h and 36~h on-source integration time, respectively.  This sets the condition to detect the polarized dust emission across extragalactic disks like the HASHTAG/DOWSING sample, but is 100$\times$ deeper than JCMT.  Luminous and ultra-luminous infrared galaxies ((U)LIRGs) are brighter than the HASHTAG/DOWSING spiral galaxies, and thus they require shorter observing times (integration time $\propto$ flux$^{-2}$) of order of a few hours. For a survey of roughly a hundred nearby galaxies including spirals and (U)LIRGs, the total on-source integration time will be about one to two thousand hours.  A wider bandwidth and/or a multi-band continuum camera will reduce the required observing time correspondingly.

%%%%%%%%%%%%%%%%%%%%%%%%%%%%%%%%%%%%%%%%%%%%%%%%%%%%%%%%%%%%%%%%%
%%%%%%%%%%%%%%%%%%%%%%%%%%%%%%%%%%%%%%%%%%%%%%%%%%%%%%%%%%%%%%%%%
%\vspace{0.5in}
%\setlength{\parskip}{6pt}
\subsection{Physics and chemistry of the interstellar medium}
\label{sec: line survey}
\subsubsection{Context and open questions}

The relationship between the cold ISM of galaxies and their star formation output is commonly studied and parametrised as the Kennicutt-Schmidt relation \citep[KS relation;][]{Schmidt1959, Kennicutt1989}. This empirical relation links the star formation rate (surface) density to the total H$\textsc{i}$ and H$_2$ gas (surface) density, through a power law with a measured index of $N_{\mathrm{KS,H\textsc{i}+H_2}} \sim 1.4$ across large samples of nearby galaxies \citep{delosReyes2019,Kennicutt2021}. This power index, however, is found to be about unity in tens of nearby spiral galaxies on resolved kpc-scales \citep{Wong2002,Leroy2008,Bigiel2008,Schruba2011,Leroy2013}, with significant variations in both slope and normalisation of the relation at GMC scales and from galaxy to galaxy \citep{Sanchez2021,Pessa2021,Pessa2022}. At high enough sub-kpc scales, the KS relation eventually breaks down, a feature that can be exploited to place constraints on important timescales in the star formation process \citep[e.g.][]{Kruijssen2014, Chevance2020Review}. 

In addition to this range of physical scales, the KS relation can be studied in multiple gas tracers, each sensitive to a different phase of the ISM. For example, it is found that it is the H$_2$ (molecular) gas rather than the H$\textsc{i}$ (atomic) gas that determines the linear KS relation, even in atomic gas-dominated regions \citep[e.g.][]{Schruba2011}. Furthermore, \citet{Gao2004} first found that the dense gas, as traced by high-dipole-moment molecules like HCN, linearly traces the star formation rate  ($N_{\mathrm{KS,dense}} \sim 1.0$), with fewer galaxy-to-galaxy variations. This may indicate that the denser phase traced by HCN is the fundamental unit where H$_2$ gas is turning into stars at a constant efficiency \citep{Gao2004,Wu2005,Gao2007,Lada2010,Lada2012,Zhang2014,Greve2014,Liu2015}. 

Despite these breakthroughs, we are far from having a full picture yet. In the past ten years, Galactic observations of an even higher-critical-density molecule, N$_2$H$^+$, and observations of HCN and HCO$^+$ in our galaxy as well as in central regions of nearby galaxies have brought into question whether some of the most popular extragalactic dense gas tracers, HCN and HCO$^+$, are indeed tracing the highest-density gas as commonly assumed \citep[e.g.][]{Kauffmann2017,Pety2017,Tafalla2023}. Differences between the spatial distributions of N$_2$H$^+$ and HCN in individual galactic giant molecular clouds (GMCs), and the abnormally enhanced HCN in nearby galaxy centers \citep{Usero2015,Bigiel2016,Gallagher2018a,Gallagher2018b,JimenezDonaire2017,JimenezDonaire2019,Jiang2020} all suggest that the chemical abundances and excitation conditions are playing crucial roles. 

Such results make it obvious that further progress requires moving away from single-line observations, in favor of deep, high resolution, multi-tracer and multi-transition observations, alongside sophisticated non-local-thermal-dynamic-equilibrium (non-LTE) excitation and radiative transfer modeling  \citep{Leroy2017,Liu2021,Leroy2022,Teng2022,Teng2023,Neumann2023,Priestley2023}.

\begin{figure}
    \centering
    \includegraphics[width=\linewidth]{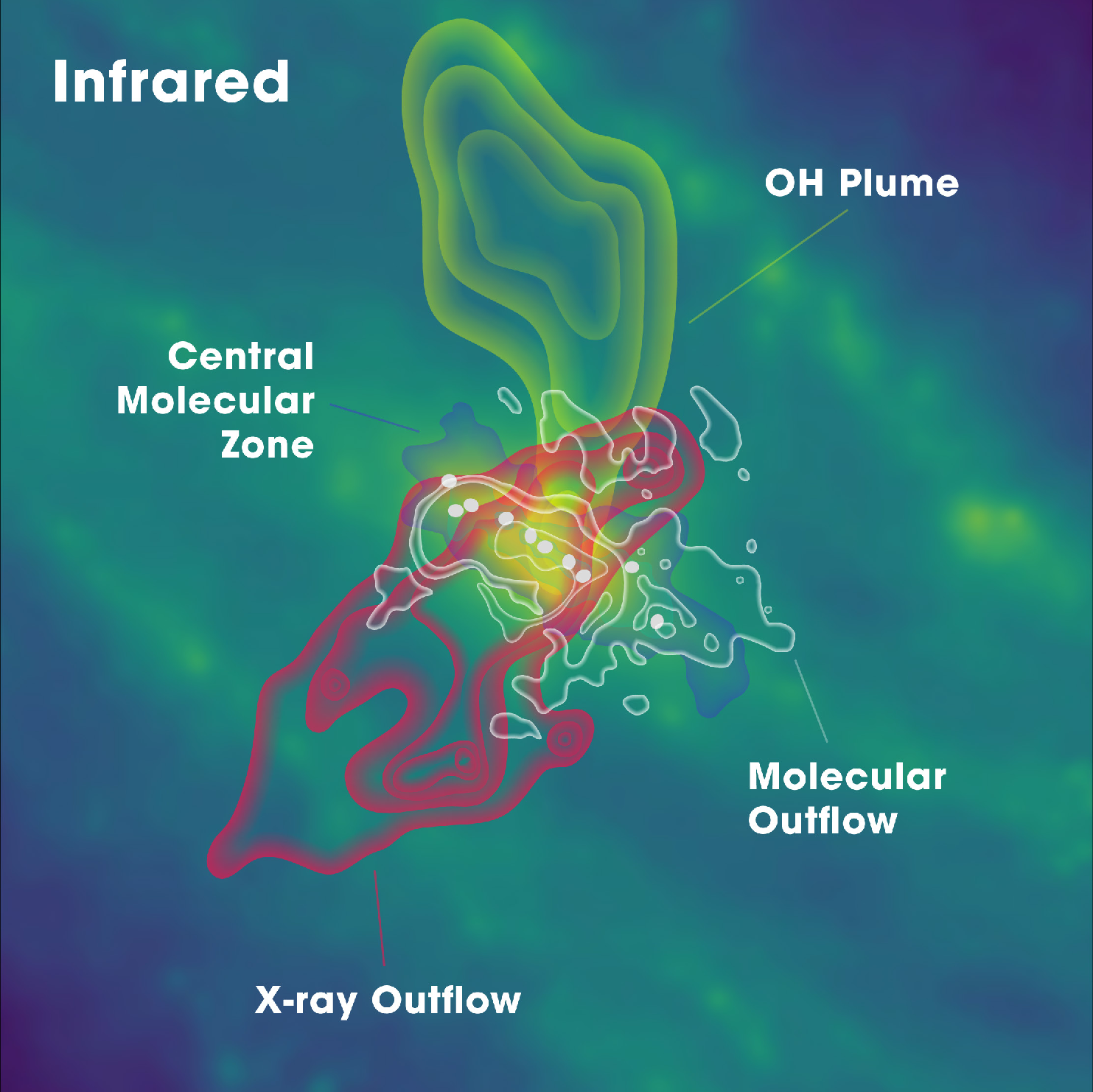}
    \caption{%
    Artistic rendering (informed by real observations) of the key physical environments in the central kilo-parsec of the nearest starburst galaxy NGC~0253 from \citet{Martin2021}.
    \label{fig:NGC253}
    }
\end{figure}

\begin{figure*}
    \centering
    \includegraphics[width=0.93\textwidth]{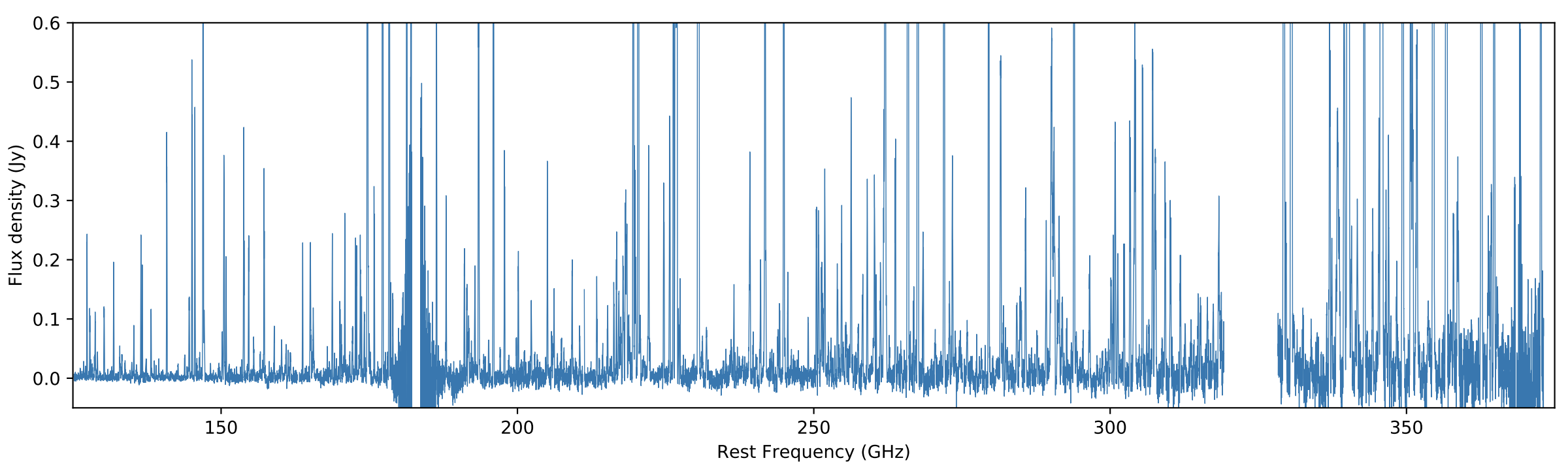}\\
    \includegraphics[width=0.7\textwidth, trim=4cm 0 0 0]{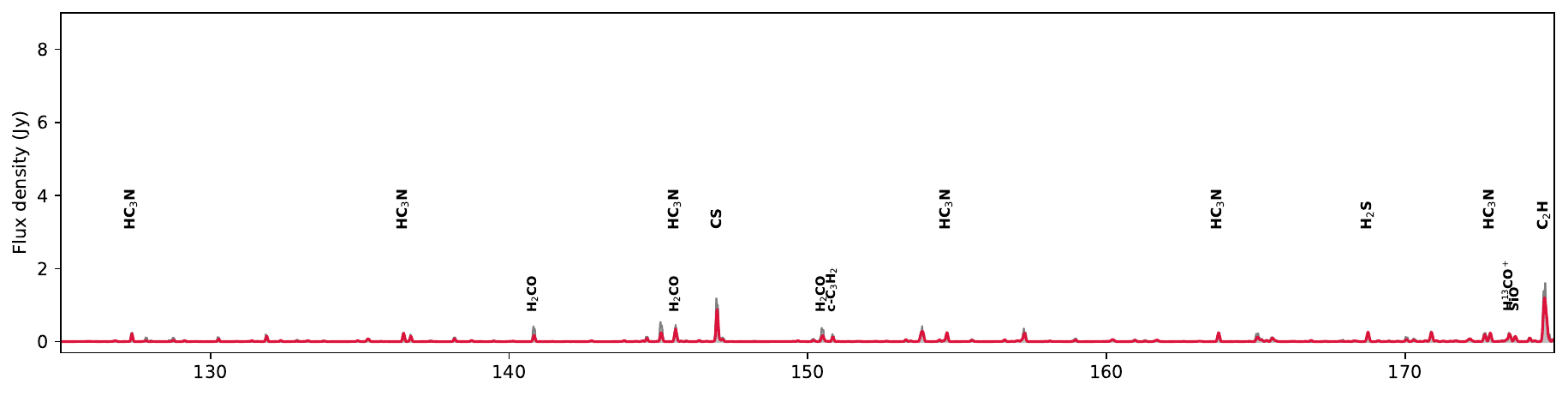}\\[-0.2cm]
    \hspace{1.5cm}%
    \includegraphics[width=0.7\textwidth, trim=4cm 0 0 0]{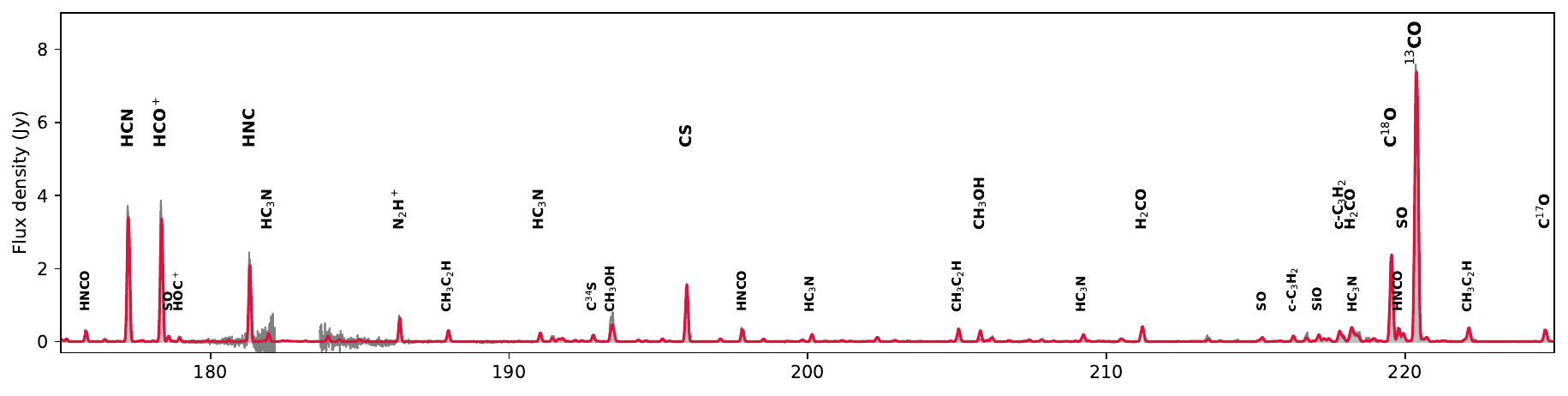}\\[-0.2cm]
    \hspace{3.0cm}%
    \includegraphics[width=0.7\textwidth, trim=4cm 0 0 0]{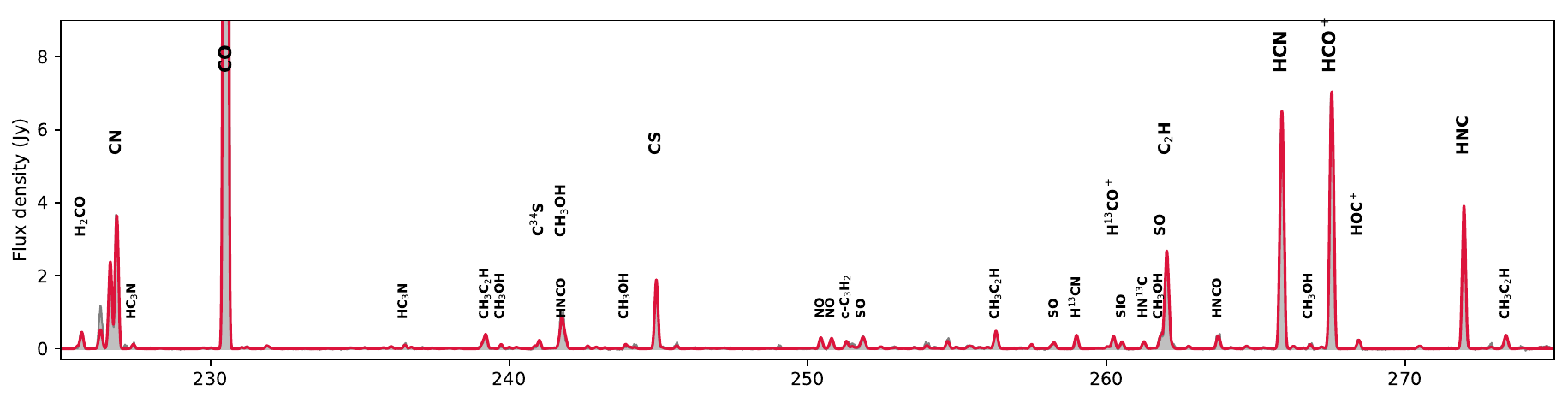}\\[-0.2cm]
    \hspace{4.5cm}%
    \includegraphics[width=0.7\textwidth, trim=4cm 0 0 0]{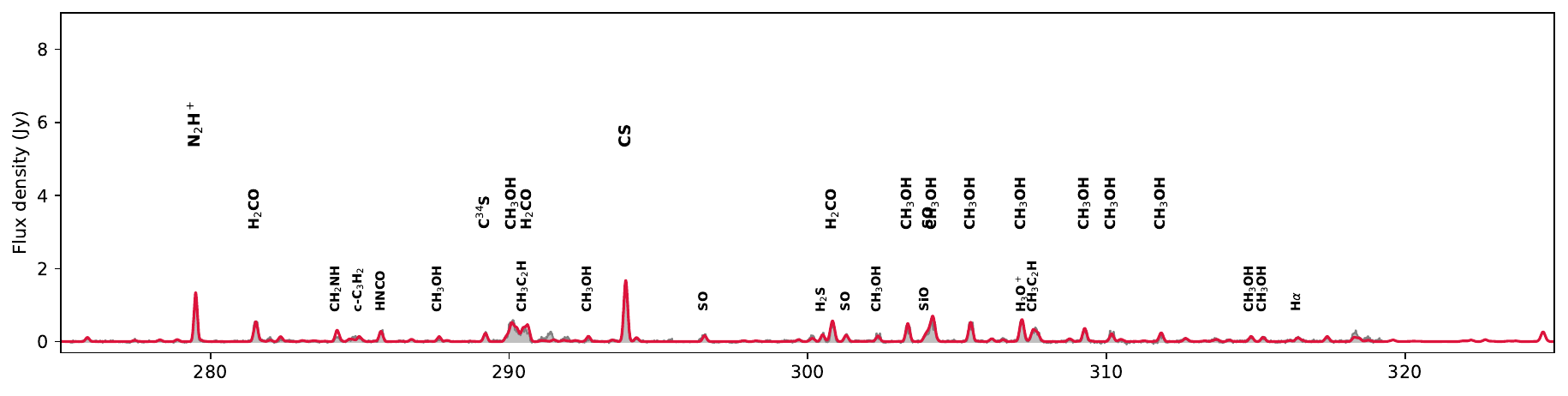}\\[-0.2cm]
    \hspace{2.5cm}%
    \includegraphics[width=0.7\textwidth, trim=0 0 4cm 0]{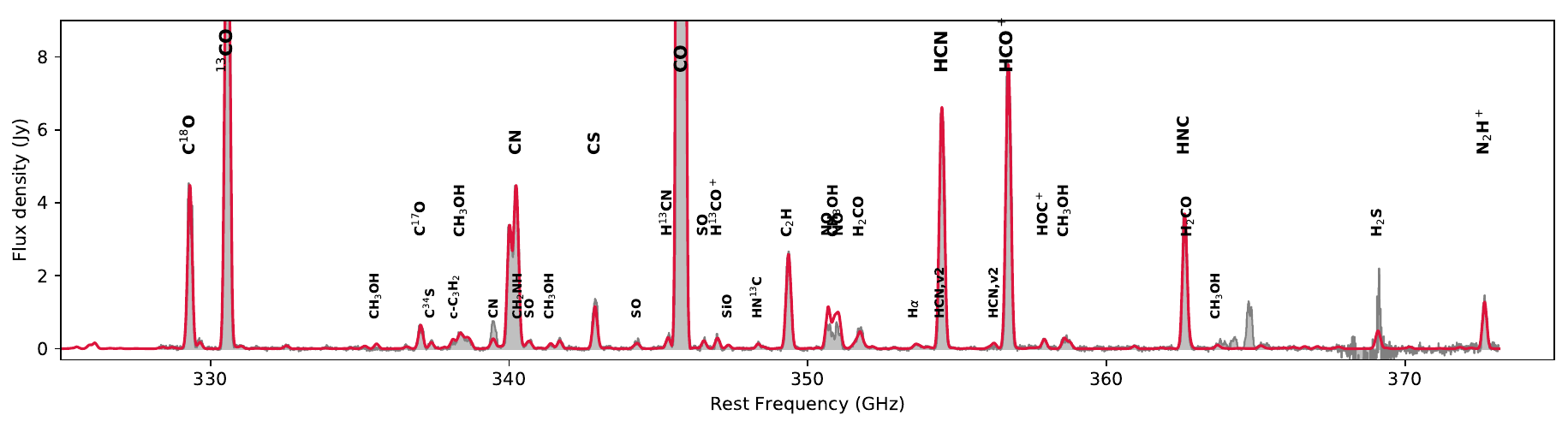}\\[-0.3cm]
    \caption{%
    Spectrum of the central kiloparsec of the nearest starburst galaxy NGC~0253 from the ALCHEMI survey using the ALMA ACA 7-m array \citep{Martin2021}.
    \label{fig:ALCHEMI}
    }
\end{figure*}

\begin{figure*}
    \centering
    \includegraphics[width=0.8\textwidth]{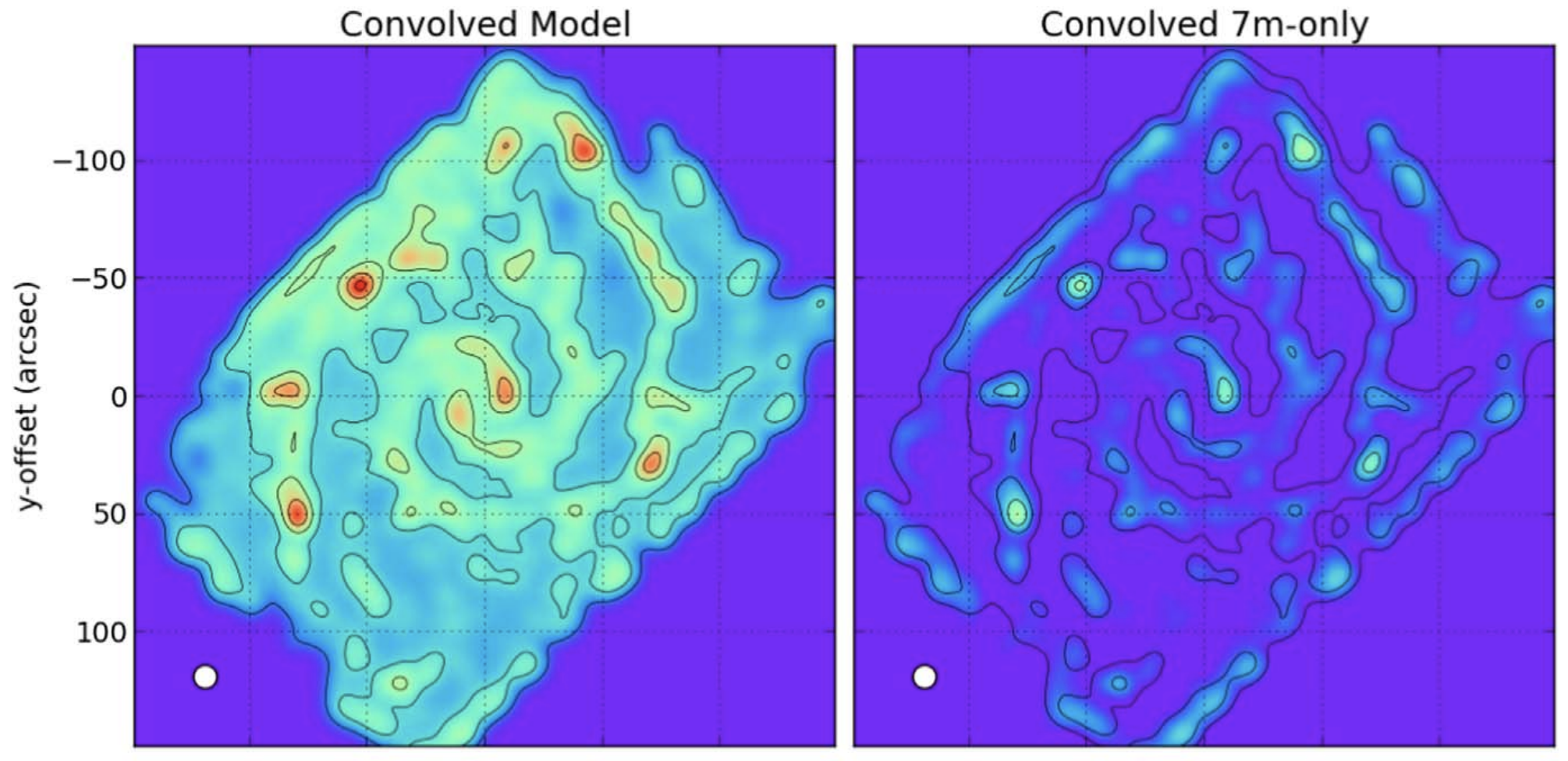}
    \caption{Figure from \cite{Leroy2021a} showing the effect of missing flux. The left panel shows a short-spacing corrected image of NGC~0628 in the CO(2-1) line, and the right panel shows a 7-m only cleaned image (i.e., without short-spacing correction using any TP data). 
    }
    \label{fig:missing-flux}
\end{figure*}

\subsubsection{The role of AtLAST in probing the physics and chemistry of the ISM}

Sensitive, large-area and high-resolution multi-species, multi-transition line observations of nearby galaxies are a critical resource for obtaining an accurate understanding of the ISM and star formation physics in nearby galaxies.  What is also needed are large surveys that provide significant statistics over a broad range of galactic environments (see example in Figure~\ref{fig:NGC253}), from low-density regions at the transition between the atomic and molecular phases, to giant star-forming complexes, and up to extreme environments such as circumnuclear regions, including gas exposed to radiative and mechanical feedback from starbursts and active galactic nuclei, and material affected by galaxy mergers. 

Such a goal is impossible to reach with current facilities. First, a high spatial resolution (physical scale of hundreds of pc or better) is required to clearly distinguish physical processes that operate on a large range of spatial scales.  In the submillimeter, where critical high-energy-level transitions of important molecules are found (see Table \ref{tab:spectral lines}), the resolution from current sub-mm single-dish telescopes (all $\leq 15$~m in size) is insufficient. Even the largest current millimeter telescope, the Large Millimeter Telescope (LMT), with an aperture of 50-m cannot come close to achieving the desired observations due to a lack of frequency coverage in the submm. 

Secondly, the largest submillimeter interferometer, ALMA, has the sensitivity to achieve incredible multi-species, multi-transition observations, however, it has two severe limitations that will not be resolved even by planned upgrades: (1) extended emission is filtered out, and (2) ALMA has a very small field of view compared to the angular sizes of nearby galaxies. For example, the ALMA Cycle 5 large program ALCHEMI has conducted a comprehensive survey of spectral lines in Bands 3, 4, 6, and 7 at $1''$ resolution in the nearest starburst galaxy, NGC~0253 \citep{Martin2021}.  Figure \ref{fig:ALCHEMI} shows how these observations reveal an astonishing number of spectral lines, from simple species all the way to complex organic molecules, but only within its central molecular zone ($850 \times 340 \; \mathrm{pc}$, see Fig. ~\ref{fig:NGC253}), leaving a vast range of other galactic environments unexplored at such depth. Another important issue is the missing flux. Despite the large collecting area and sensitivity of ALMA's main array, the total power (TP) antennas are far from being sensitive enough for scanning faint lines in a large sample of galaxies. 
Without adequate single-dish data, it has been demonstrated that missing flux can be severe for faint lines \citep[e.g.,][]{Leroy2021a,Liu2023c}, especially missing line flux at GMC edges and in inter-arm regions by more than 50\% thus propagating to a huge error in line ratios. Figure~\ref{fig:missing-flux} shows an example of missing extended flux when mapping a nearby galaxy with ALMA. In conclusion, because of the limitations listed above, ALMA is not the right instrument to expand a survey like ALCHEMI to many more galaxies (to reveal galaxy-galaxy variations), as it cannot recover the multi-line physics accurately at large scales.

AtLAST will uniquely enable such surveys of the multi-species, multi-transition molecular/atomic lines in hundreds of nearby galaxies, representing all kinds of galactic environments and thus physical mechanisms, at GMC/star-forming complex scales. ALMA will be highly complementary to achieving sub-cloud resolution in a limited number of targeted regions of interest. We summarize below some of the key science questions that can only be answered with AtLAST. 

\paragraph{Is the gas-star formation relation universal?} 
This question is fundamental in our understanding of star formation. The HERACLES \citep{Leroy2009}, PHANGS \citep{Leroy2021b} and EDGE-CALIFA \citep{Bolatto2017,Sanchez2021} surveys have achieved sub-kpc scale studies of the KS relation in a single $^{12}$C$^{16}$O line in about a hundred nearby star-forming galaxies representing the local star-forming main sequence. The EMPIRE \citep{JimenezDonaire2019} and MALATANG \citep{Tan2018,Jiang2020} surveys have achieved $\ge$~kpc-scale dense-gas KS relation studies in about ten IR-bright galaxies. However, the debate about the emissivity of dense gas tracers and the scatter in star formation efficiency at low densities are strongly limiting our theoretical understanding of the star formation processes. Is there a density threshold? Is local gas pressure playing a role? What drives the star formation efficiency / timescale variations? How well can we trust the surface mass densities from CO or HCN, using conversion factors? These are all related questions that need AtLAST's sensitivity, GMC-scale resolution, mapping speed and large field of view. 

\paragraph{What are the gas densities and temperatures?} 
Multi-transition non-LTE modeling of molecular/atomic lines is the most powerful tool to measure gas densities, temperatures and line excitation conditions for the multi-phase gas in galaxies. Fifteen years ago, only very few galaxies had ground-based (sub)millimeter observations of the CO spectral line energy distribution (SLED) up to $J=6\text{--}5$. The \textit{Herschel} space observatory opened a new window to studying the CO SLEDs in over 160+ nearby galaxies (\cite{Greve2014,Kamenetzky2014,Rosenberg2015,Liu2015,Kamenetzky2016,Kamenetzky2017,Lu2017,Kamenetzky2018}), however, with $\sim 30 \text{--} 40''$ resolution these observations do not resolve external GMCs. Moreover, the limited spectral resolution of the instruments on board the {\it Herschel} satellite prevents us from using these data to study spectral line profile variations in different H$_2$ tracers and CO transitions. The latter is a promising avenue for understanding the relation between extreme H$_2$ gas excitation and feedback mechanisms at play in galaxies, as shown by \cite{MontoyaArroyave2023} and \cite{MontoyaArroyave2024} using sensitive multi-J CO and [CI] APEX observations of tens of local (U)LIRGs. 

High-resolution, multi-transition, multi-species non-LTE studies have only been made very recently in a few IR-bright galaxies using ALMA (e.g., CO isotopologues $J=1\text{--}0$ to 3--2 \cite{Teng2022,Teng2023}; CO $J=1\text{--}0$ to 4--3 + [C\textsc{i}] ${}^{3}P_{1}\text{--}{}^{3}P_{0}$ \cite{Liu2023b,Liu2023c}). Also with ALMA, the PHANGS survey has mapped of order one hundred nearby galaxies in CO $J=2\text{--}1$, but obtaining similar coverage of additional lines, e.g., CO $J=3\text{--}2$, is very difficult and time consuming. The ALMA studies are limited to a few galaxy centers, or a few off-center regions (e.g., like the GMC study in M83's XUV disk; \citealt{Koda2022}), which do not probe the variety of galactic environments. Simulations and theory have shown that the emissivity of different species can vary significant with galactic environments, thus the conversion factors do vary accordingly. The gap from single-line conversion factor-based studies to multi-line non-LTE studies has to be filled in order to understand how gas density and temperature play a role across the star-forming disks/environments, and to anchor simulations and theories (e.g., photodissociation region, cosmic ray dominated region, X-ray dominated region and shock physics \citep{Kaufman1999,Kaufman2006,Papadopoulos2010,Papadopoulos2011,Bisbas2015,Bisbas2017,Meijerink2005,Meijerink2007,Wolfire2022}.

\subsubsection{AtLAST surveys of multiple gas/dust tracers in the nearby Universe}

A complete survey of multiple (sub)millimetre lines in a large number of galaxies at hundreds of parsec scales will enable unique studies of gas and chemistry in galaxies. 
The brightest (sub)millimetre lines are listed in Table~\ref{tab:spectral lines}, but many fainter lines from species like HC$_3$N, HC$_3$HO, HOCN, HOCO$^+$ \citep{Martin2021} are also expected. 
The sample of nearby galaxies should include a variety of southern nearby galaxies. For main-sequence types, there are $\sim 100$ galaxies from the PHANGS-ALMA large program which have full-disk CO(2--1) line mapping at $\sim 1''$ (about 30--300~pc). This survey (the large program and a few pilot programs) used about a hundred hours the ALMA main array and 600+ ALMA ACA including the total power observations. The $^{13}$CO(2--1) isotopologue line is observed simultaneously, but due to the limited sensitivity to the extended emission, only brightest clouds are imaged in high quality. 
AtLAST multi-line mapping of these galaxies will be an unprecedented legacy survey at moderately high resolution ($\sim 2$--$4''$) starting from Band 7, or the CO(3--2) transition. The 3--2 transitions of CO isotopologues have been found crucially useful in determining the CO-to-H$_2$ conversion factors \citep{Teng2022,Teng2023}. The Band 8 CO(4--3) and [C~{\sc i}] lines are of great importance in constraining gas density and even possibly stellar and AGN feedback at resolved scales \citep{Saito2020,Liu2023c,Liu2023b}. An ALMA ACA survey of the CO(4--3) and [C~{\sc i}](1--0) lines was conducted in 36 (ultra-)luminous galaxies at $\sim 3''$ \citep{Michiyama2021}, leading to the finding of unusually low-[C~{\sc i}] galaxies which are still puzzling. However, the deep, multi-tracer single-dish APEX observations of local (U)LIRGs presented by \cite{MontoyaArroyave2023} include [CI](1-0) coverage for 17 sources (some of which in common with the ACA sample of \cite{Michiyama2021}) and show a significantly higher [CI] flux recovered by APEX, resulting in an average [CI](1-0)/CO(1-0) luminosity ratio as high as $0.210\pm0.009$ in local (U)LIRGs.  
The AtLAST survey of lines will be highly complementary to ALMA main array and ACA surveys which either have much higher resolution (and usually missing fluxes or would require hundreds of hours total power observing time) or only have the sensitivity to observe the most infrared-luminous starbursts.

\begin{figure*}[t]
    \centering
    \includegraphics[width=0.9\textwidth]{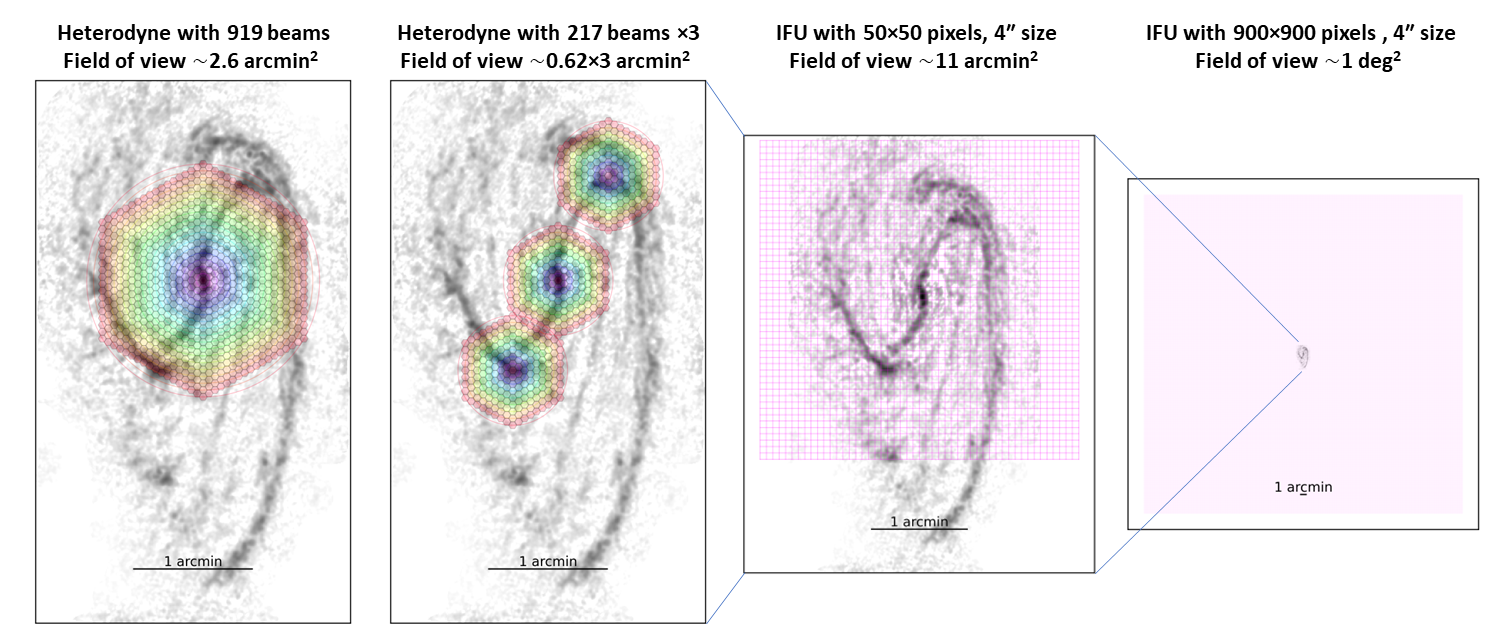}
    \caption{%
    Footprints of various assumed instruments (from left to right): a 919-beam (18-layer) heterodyne receiver, a 217-beam (9-layer) heterodyne receiver, an IFU with $50 \times 50$ pixels, and an IFU with $900 \times 900$ pixels, on top of the nearby galaxy NGC~3627's CO(2--1) emission \citep{Leroy2021b}. 
    The heterodyne receiver's hexagon units and the IFU pixels are assumed to have a spacing/size of $4''$, which is about the AtLAST 50-m angular resolutions at Band 7--10, and is about half of the angular resolutions at Band 4--5. 
    A large-format IFU as shown in the last panel can most efficiently observe multiple galaxies, with each galaxy fully sampled at resolutions as shown in the first three panels. 
    \label{fig:heterodyne}
    }
\end{figure*}

We can consider a sample of 100 nearby galaxies with a variety of types:
\begin{itemize}[topsep=0pt,itemsep=0ex,partopsep=0ex,parsep=0ex]
    \item main-sequence spirals
    \item starbursts
    \item AGNs/composite galaxies
    \item dwarfs
\end{itemize}
and we observe the Band 6 (CO 2--1 isotopologue), Band 7 (CO 3--2), Band 8 (CO 4--3 and [C~{\sc i}] 1--0), Band 9 (CO 6--5), and Band 10 (CO 7--6 and [C~{\sc i}] 2--1) lines. 
In each Band, we assume that the receiver has sufficient bandwidth to cover all the key lines simultaneously, that is, $\sim 16$--20~GHz per sideband for a double-sideband receiver. We also require a multi-beam receiver or integral field unit (IFU) to efficiently map the galaxies. In the case of a multi-beam heterodyne receiver, assuming a 919-beam (18-layer) array, as shown in Fig.~\ref{fig:heterodyne}, a galaxy like NGC~3627 at a distance of $\sim 11$~Mpc will be covered by roughly ten pointings. For slightly more distant local galaxies like those in the EDGE-CALIFA or MaNGA samples, fewer pointings are needed. In the case of an IFU, the field of view is much larger thus usually these nearby galaxies can be covered in one pointing, but the spectral resolution is much lower, of only $\sim 300$--3000~km/s ($R \sim 100\text{--}1000$). Our survey of line ISM physics do not strongly require resolving the lines, but require accurate integrated line fluxes and ratios, therefore an IFU with a spectral resolution $R \gtrsim 1000$ will be more efficient than a multi-beam receiver (although a multi-beam receiver will enable kinematic studies thus can add additional value). 

Then, we estimate the line fluxes and sensitivity requirements. Taking NGC~3627 as an example (Fig.~\ref{fig:heterodyne}), its CO(2--1) surface brightness ranges from $\gtrsim 1000\,\mathrm{K\,km\,s^{-1}}$ to $\sim 30\,\mathrm{K\,km\,s^{-1}}$ at $\sim 8$~kpc away from the center along the spiral arm. The peak temperature ranges from $\sim 8$~K to $\sim 1$~K from the center to the $\sim 8$~kpc arm. The $^{13}$CO lines are in general $\sim 10 \times$ fainter than the $^{12}$CO lines \citep[e.g.,][]{Wilson1994,Taniguchi1998,Henkel2014,JimenezDonaire2017,Cormier2018,Cao2023}, and the [C~{\sc i}](1--0) lines are generally $\sim 5\text{--}10\times$ fainter than $^{12}$CO(2--1) \citep{Liu2023c}. The ratios of CO(1--0) line fluxes to those of HCN and HCO$^{+}$ 1--0 are typically $\sim 30 \text{--} 100$ at resolved kpc scales \citep{Bigiel2016,Gallagher2018a,Gallagher2018b,JimenezDonaire2019,Neumann2023}. Thus, in general we will need to reach an RMS of $\sim 0.1\,\mathrm{K\,km\,s^{-1}}$ to achieve S/N~$\gtrsim 3$ detections for all the aforementioned lines. Given a line width of $\sim 30\,\mathrm{km\,s^{-1}}$ and the AtLAST beam of $2''$--$7''$ from Band 10 to Band 6 ($\mathrm{Jy/K} \sim 1.5 \text{--} 2$), this RMS corresponds to $\sim 5 \text{--} 7$~mJy at these bands. 

Using the \href{https://www.atlast.uio.no/sensitivity-calculator/}{AtLAST sensitivity calculator}, we compute that at an elevation of 45~deg, an observing frequency 230/345/461/492/691/809~GHz, with a bandwidth of 0.023/0.035/0.05/0.05/0.07/0.08~GHz ($\sim 30\,\mathrm{km\,s^{-1}}$), and an H$_2$O profile percentile 80/80/50/50/20/20, the on-source integration time will be $\sim 1$/10/24/84/25/60 minutes. Thus the expected typical on-source observing time is only $\sim 3.5$ hours for all the major Band 6--10 lines in one pointing in a galaxy like NGC~3627. The full survey of $\sim 100$ nearby galaxies with one IFU pointing (or multiple pointings with multi-beam heterodyne receiver) for each galaxy will be at scales of hundreds to a thousand hour.

%%%%%%%%%%%%%%%%%%%%%%%%%%%%%%%%%%%%%%%%%%%%%%%%%%%%%%%%%%%%%%%%%
%%%%%%%%%%%%%%%%%%%%%%%%%%%%%%%%%%%%%%%%%%%%%%%%%%%%%%%%%%%%%%%%%
%\vspace{0.5in}
%\setlength{\parskip}{6pt}
\subsection{An ``SDSS-like" submillimeter survey for galaxy evolution studies in the nearby Universe}
\subsubsection{Context and open questions}

The evolution of galaxies is a fundamental question and a complex process in astrophysics. The amount of cold gas, the rate of star formation, and the environment of galaxies all play a critical role in shaping galaxy evolution. How the cold gas is accreted into galaxy systems, how the star formation is set by the turbulent ISM properties, and how the gas and dust is influenced by the external gravitational forces in dense environments are all key processes demanding detailed understanding with large-sample surveys. 
While millions of galaxies have been observed in optical spectroscopic surveys such as SDSS and more recently DESI, only hundreds of them have constraints on the molecular gas reservoir --- thus leading to a giant gap in our knowledge of galaxy evolution \citep{Saintonge2011,Saintonge2017,Saintonge2022,Tacconi2018,Liu2019b,Tacconi2020}. 
An extra-large submillimeter survey in the local Universe will be a substantial breakthrough. It will help bridge the gap with optical, near-infrared (e.g., \textit{Euclid}, LSST, \textit{Roman}), and radio (e.g., with the SKA) surveys, which would otherwise become even greater. Extending galaxy samples with molecular gas measurements by orders of magnitude will lead to the most important statistical understanding of cold gas and star formation, and environments and galaxy quenching. Possibly tens of thousands of molecular line detections could be achieved with an ``SDSS-like'' survey with AtLAST, highly complementary to the atomic gas surveys with the future SKA. 

The statistical properties of the molecular ISM gas in low-mass ($M_{\star} \lesssim 10^{9}\,\mathrm{M_{\odot}}$) galaxies is one of the major unknowns in galaxy surveys. Low-mass galaxies have sub-solar metallicities and correspondingly low CO abundance compared to the H$_2$. With less dust shielding, radiation fields created by the young massive stars are in general stronger and harder. Many low-mass galaxies are also satellites of massive galaxies, e.g., the LMC and SMC as described in Sec.~\ref{sec: MCs}. 
The xCOLDGASS survey \citep{Saintonge2011,Saintonge2017} is the largest millimetre CO survey of local galaxies with $\sim 950$~h observing time with the IRAM 30~m telescope. This survey has been able to constrain the local-Universe CO luminosity distribution down to $10^{7.5} \, \mathrm{K\,km\,s^{-1}\,pc^{2}}$ (or a stellar mass limit of $10^{9.0}\,\mathrm{M_{\odot}}$). The CO luminosity function (and the consequent H$_2$ mass function) in the local universe is a important tool in constraining models \citep[e.g.][]{Vallini2016} and anchoring studies of the redshift evolution of gas reservoirs \citep[e.g.,][]{Liu2019b,DessaugesZavadsky2020,walter20}. The scaling relations between molecular gas mass and other global galaxy properties (stellar mass, SFR, morphology,...) are also proving to be very constraining for hydrodynamical simulations \citep[e.g.][]{dave20}. Despite all these successes, with only $\sim1000$ galaxies, combining xCOLDGASS and other similar surveys \citep{boselli14_1, cicone17, colombo20, Wylezalek2022}, we can only scratch the surface, as significantly more statistics are required to disentangle the competing effects of different evolutionary mechanisms. It is particularly crucial to improve statistical constraints of the H$_2$ content of low-$M_{*}$ galaxies, not only because these are the most common galaxies in the Universe, but also because including these sources allows us to expand the dynamic range of current scaling relations and so provide more reliable constraints for theoretical models, as shown by pilot studies using $\sim330$ hours of APEX observations (ALLSMOG survey: \citealt{cicone17, Hagedorn2024}).

\begin{figure*}
    \centering
    \includegraphics[width=0.99\linewidth]{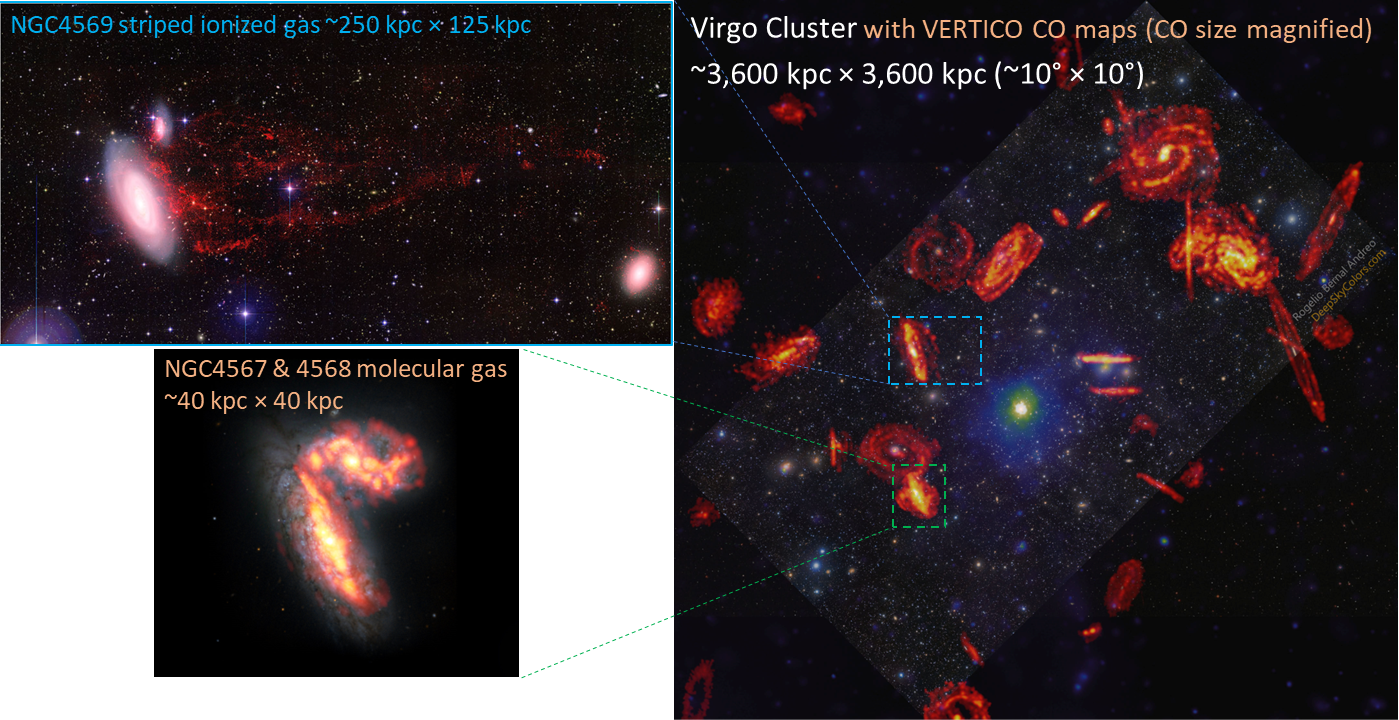}
    \caption{%
    The right panel shows the Virgo cluster's optical and CO images from the VERTICO ALMA large program \citep[][\href{https://www.verticosurvey.com/}{see the website}]{Brown2021} and the \href{https://apod.nasa.gov/apod/ap150804.html}{Astronomy Picture of the Day}. The CO maps of each individual member galaxies are scaled up for visualization. 
    The top left panel is adapted from \cite{Boselli2016a} and \cite{Boselli2022}, showing the very extended stripped materials from the Virgo cluster member galaxy NGC~4569. The image size is $50' \times 25'$ or $250 \mathrm{kpc} \times 125 \mathrm{kpc}$. The red image shows the stripped ionised gas extending up to 150~kpc. 
    The bottom left panel is adapted from \cite{Brown2021}, showing the cluster members NGC~4567 and NGC~4568. 
    The Virgo cluster extends at least $10^{\circ} \times 10^{\circ}$ in the sky. 
    }
    \label{fig:NGC4569 stripped materials}
\end{figure*}

Due to these sample size limitations, current surveys are unable to accurately quantify the impact of large-scale environment and the position of a galaxy with respect to the cosmic web on the molecular ISM. We know from HI observations that galaxies in dense regions (such as clusters) generally have a reduced atomic gas content and star formation activity, compared to their counterparts in low density environments \citep[e.g.][]{Haynes86}. For example, Fig.~\ref{fig:NGC4569 stripped materials} shows an example of stripping of HI gas in a galaxy infalling into the Virgo cluster. A range of mechanisms, from gravitational interactions to hydrodynamic processes, can explain the impact of the dense cluster environment on the HI contents of galaxies.  What is still not totally understood is their effect on the molecular gas component, which is much more embedded into the gravitational potential of the disc and mainly located within giant molecular clouds thus much less sensitive to  external perturbations than the atomic phase \citep[e.g.,][]{Boselli2014a,Boselli2022}.

An extra-large survey of $>10,000$ local galaxies to measure accurate molecular gas masses, especially if extending the coverage of current surveys into the dwarf galaxy regime ($M_{\ast} < 10^{9.0}\,\mathrm{M_{\odot}}$) and covering the full range of extragalactic environments, from field to clusters, is the only approach to achieve the statistics required to address several important questions in galaxy evolution, such as those described below.

\subsubsection{The role of AtLAST in the study of the local galaxy population}

AtLAST is ideally suited to shed light on these important topics. Thanks to its high sensitivity and large field of view, it can be used to map in a blind survey a sufficiently large fraction of the nearby sky to measure with unprecedented sensitivity and statistics the molecular gas mass function of galaxies in the local Universe for objects located in different density regions, from voids and filaments to groups and massive clusters. 

\paragraph{What is the abundance of molecular gas in the local Universe and its distribution?} The molecular gas mass function is a very useful constraint to semi-analytic models and cosmological simulations \citep[e.g.][]{lagos15,diemer19,dave20}. Unlike for atomic gas, where large blind HI surveys have allowed for accurate measurement of the HI mass function in the local Universe \citep{Jones2018}, we do not yet have a deep and complete census of molecular gas in the nearby Universe.  With current instruments, large CO blind surveys are impossible, and therefore all the $z=0$ molecular gas mass functions are based on optically- or IR-selected samples \citep[e.g.][]{keres03, andreani20, fletcher21}. AtLAST is the only facility which would enable a blind survey for molecular gas in the nearby Universe, which is essential to anchor the all important studies of the redshift evolution of the baryonic mass of galaxies and their star formation output \citep[e.g.][]{walter20}.

\paragraph{Molecular gas scaling relations in the dwarf galaxy regime} The first large systematic molecular gas surveys of the past 15 years, both at low and high redshifts, have been transformational in our understanding of the central roles of the baryon cycle and star formation efficiency in the galaxy evolution process.  With these surveys, we now have a good overview of the basic scaling relations between total molecular gas mass and other galaxy observables, but only in relatively massive galaxies ($M_{\ast}>10^{9.5}M_{\odot}$).  Systematic surveys for CO in lower mass galaxies \citep[e.g. ALLSMOG,][]{cicone17} are very challenging with current instrumentation, resulting in low detection rates. With the increased sensitivity of AtLAST and improved survey strategies enabled by its very large field of view (see Sec. \ref{sec:largeCOsurveys}), it will be possible to expand molecular gas scaling relations well into the regime of dwarf galaxies ($M_{\ast}<10^{9}M_{\odot}$), either through direct line detection or stacking of the many undetected objects that would be covered by a blind CO line survey.

\paragraph{Why do galaxies stop forming stars?} This question, often referred to as "star formation quenching", is central to our understanding of galaxy growth, and as of yet remains unanswered.  A galaxy can stop forming stars if it runs out of cold gas, or if any cold gas it has is somehow prevented from forming stars. All the physical mechanisms that are invoked to explain quenching work  by either affecting the gas contents of the galaxies, or the star formation efficiency out of any available gas.  Therefore, only by having direct information about the cold atomic and molecular gas mass of galaxies can we accurately pinpoint the mechanisms that are causing star formation activity to shut down. Current evidence suggests that both a reduction of gas reservoirs and a decrease in star formation efficiency are at play \citep[e.g.][]{colombo20, Saintonge2022}, but the sample size limitations are currently preventing us from understanding why these changes are happening.  AtLAST will enable the assembly of a sample of $\sim 10^5$ galaxies with CO detections (two orders of magnitude more than current samples, see Sec. \ref{sec:largeCOsurveys}) which will finally allow us to disentangle the complex physical processes behind star formation quenching.

\subsubsection{AtLAST surveys to probe star formation and galaxy evolution in the nearby Universe}
\label{sec:largeCOsurveys}

The primary goal of this survey is to detect CO lines in a significantly large sample that is a factor of $> 10 \text{--} 100 \times$ larger than any existing survey in the local Universe. This consists of two types of targets: a mass-complete sample of field galaxies in the local Universe (e.g., redshift $z \sim 0.01 \text{--} 0.38$, or distances $\lesssim 2,000$~Mpc); and a sample of (super)clusters for the dense environments. 

\begin{figure}[t]
    \centering
    \includegraphics[width=\linewidth]{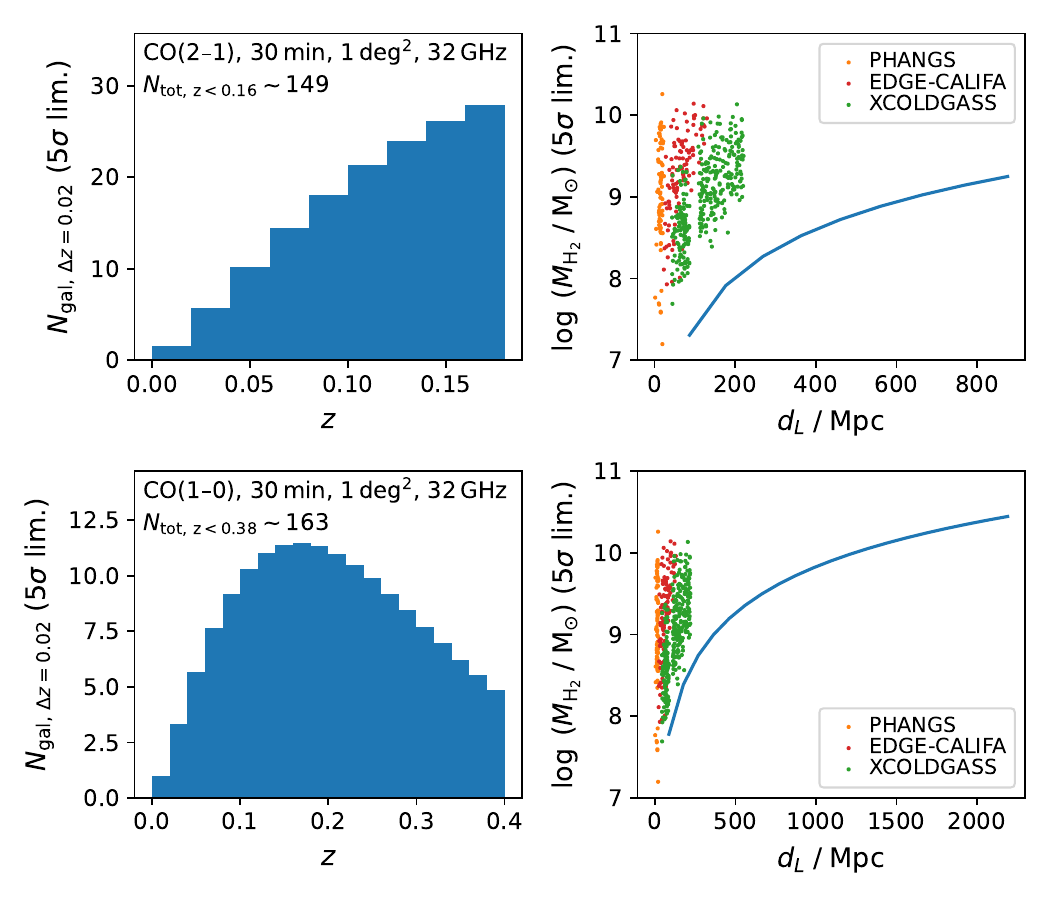}
    \caption{%
    Predicted numbers of CO emitters in an 1~deg$^2$ field of view for 30~min on-source integration with the Band~3 CO(1--0) line (\textbf{lower panels}) and with the Band~6 CO(2--1) line (\textbf{upper panels}). With a bandwidth of 32~GHz, the CO(1--0) line frequencies are $\sim 115$--83~GHz, corresponding to $z \sim 0$--0.38, and for CO(2--1) line the frequencies are $\sim 230$--198~GHz, corresponding to $z \sim 0$--0.16. The numbers of CO emitters are calculated by integrating the CO luminosity distribution in \citep[][Fig.~6]{Saintonge2017} from the 5$\sigma$ detection limit $L^{\prime}_{\mathrm{CO}}$ to $10^{11} \; \mathrm{K\,km\,s^{-1}\,pc^{2}}$. \textbf{Left panels} are the predicted numbers of CO emitters per redshift bin of $\Delta z = 0.02$ for CO(2--1) and $\Delta z = 0.01$ for CO(1--0). The total predicted CO detections above 5$\sigma$ are $\sim 150$--160 for the CO(2--1) and CO(1--0) lines. \textbf{Right panels} are the molecular hydrogen mass $M_{\mathrm{H_2}}$ detection limit computed from the 5$\sigma$ $L^{\prime}_{\mathrm{CO}}$ detection limit using a metallicity-dependent $\alpha_{\mathrm{CO}}$ conversion factor following \citet{Bolatto2013}. A half-solar metallicity is assumed for the $M_{\mathrm{H_2}}$ limits, and a CO(2--1)/(1--0) brightness temperature ratio of 0.8 is assumed for the excitation. 
    }
    \label{fig:submm-ifu-survey}
\end{figure}

The field targets will naturally be drawn from on-going and future large optical/near-infrared and radio H~{\sc i} surveys, for example, the Dark Energy Survey (DES) and the LSST in the southern hemisphere. The sample selection will follow the COLDGASS and xCOLDGASS strategy but have great improvements in the volume completeness. The COLDGASS and xCOLDGASS IRAM 30-m large programs randomly trim a large, mass-complete, H~{\sc i}-detected sample from the GALEX Arecibo SDSS Survey \citep[GASS;][]{GASS1, GASS8} down to a few hundred galaxies for the CO observations. The final combined sample is 532 galaxies with a total of $\sim 950$~hr of IRAM 30-m observing time. The receiver has a single beam thus the field of view of each IRAM 30-m observation is limited to $\sim 22''$ at 115~GHz and $\sim 11''$ at 230~GHz. 
With a large-format submillimeter IFU taking advantage of AtLAST's very large field of view, surveys on unprecedented scales will become possible. Assuming that the IFU on AtLAST can have 1~deg$^2$ field of view and 32~GHz bandwidth, with a spectral resolution of $R \sim 1000$ or better ($\sim 40$~km/s at 230~GHz), and the spaxel units are separated by $4''$ thus having a Nyquist sampling of the $\sim 7''$ angular resolution at 230~GHz (i.e., a pixel array of $900 \times 900$ and channel number $\sim 1000$). Then, each targeted observation at a selected galaxy will have an 1~deg$^2$ field of view to allow for detections of CO line emitters in the vicinity. 
According to our best constraint of the CO luminosity function so far \citep{Saintonge2017, fletcher21}, we predict that with {30~min on-source integration} of the IFU, the number of CO line emitters will be as large as ${\sim 150}$ in a single 1~deg$^2$ field of view as shown in Fig.~\ref{fig:submm-ifu-survey}. Observing the CO(2--1) line will achieve more galaxies per redshift bin and a much lower $M_{\mathrm{H_2}}$ mass limit, thus is better than observing the CO(1--0) line for studying the faintest galaxies. 
If we observe a southern xCOLDGASS-like mass-complete sample of $\sim 500$ primary targets, each with 30~min on-source integration with the IFU, then we will eventually obtain a combined sample of $500 \times 150 \sim 75,000$ galaxies at $0 < z < 0.16$ with $>5\sigma$ CO(2--1) detections. The exact number depends on the actual CO luminosity distribution to be measured and may vary by a factor of a few, but this is truly transformational compared to any existing CO surveys. 

In the dense environments, our (super)cluster targets should contain a statistically significant number of galaxies in the densest parts of our local Universe such as those located in the Coma/A1367 supercluster (e.g. Gavazzi et al. 2010). This particular region of the sky includes $\sim 4000$ objects down to the SDSS $r$-band mag 17.7 at a typical distance of 70--100 Mpc over $\sim$ 400 square degrees, with galaxies located in two rich clusters (Coma and A1367), in several groups and filaments, and in voids. It is in the northern hemisphere though (Dec~$\sim +19$). Another widely-studied cluster, the Virgo cluster, is also in the north (Dec~$\sim +12$), spreading over 100--200~deg$^2$ in the sky with over 1300 member galaxies. 
To facilitate the observation at the AtLAST site, southern (super)clusters can be selected from the \citet{Abell1989} catalog, where over 4000 clusters are listed and each has at least 30 member galaxies. The AtLAST's unprecedented sensitivity to the extended emission will make the survey of nearby galaxy clusters not only for counting galaxies number densities and luminosity function, but also for mapping any possible stripped gas at tens of kpc scales (Fig.~\ref{fig:NGC4569 stripped materials}). If a gas-rich satellite galaxy has 50\% of its gas stripped, then an extended gas structure of a mass $M_{\mathrm{gas}} \sim 10^{8} \mathrm{M_{\odot}}$ will be spread over tens of beams. Our detection limit of 30~min on-source integration is already $\sim 2 \times 10^{7} \; \mathrm{M_{\odot}}$ (for CO-traced H$_2$ gas at $0.5\,\mathrm{Z_{\odot}}$) out to $\sim 100$~Mpc, thus it is very promising for AtLAST to open a new era for studying the extended stripped gas structures in clusters. These cluster observations can be much deeper than the field galaxy observations, as they require fewer pointings.

Combining the field and cluster CO surveys, this will provide us with an unrivaled sample of $10^{5}$ CO emitters in the local Universe, enabling the transformative science depicted above. Such a survey will require 500 hours for the field galaxies and another 500 hours for the clusters, including calibration and telescope overheads, covering $\sim 500$--1000~deg$^2$ areas. Since the observations are done in Band~6, the observing time is not very sensitive to weather conditions and therefore such a survey could make good use of time in poor weather conditions. In the calculation for this survey, we assumed a water vapour percentile of 85 and 95 for the Band~6 and Band~3 sensitivity estimates using the \href{https://www.atlast.uio.no/sensitivity-calculator/}{AtLAST sensitivity calculator}. We also assumed line widths of 200~km/s, which is 0.16~GHz and 0.073~GHz at Band~6 and Band~3, respectively.

%%%%%%%%%%%%%%%%%%%%%%%%%%%%%%%%%%%%%%%%%%%%%%%%%%%%%%%%%%%%%%%%%
%%%%%%%%%%%%%%%%%%%%%%%%%%%%%%%%%%%%%%%%%%%%%%%%%%%%%%%%%%%%%%%%%
%%   TECHNICAL CASE 
%%%%%%%%%%%%%%%%%%%%%%%%%%%%%%%%%%%%%%%%%%%%%%%%%%%%%%%%%%%%%%%%%
%%%%%%%%%%%%%%%%%%%%%%%%%%%%%%%%%%%%%%%%%%%%%%%%%%%%%%%%%%%%%%%%%
%\vspace{0.5in}
%\setlength{\parskip}{6pt}
\section{Technical requirements for AtLAST}
\label{TechnicalCase}

\subsection{Multi-band dust continuum with polarimeter}
A continuum camera with polarimeter is a key requirement to achieve the science objectives. While not essential, a multi-chroic camera would significantly increase the efficiency of observations and therefore the sample sizes we can assemble in a fixed survey time. 

The bandwidth of the continuum camera is a key parameter for determining the sensitivity of an observation, with wider bandwidths preferred. For the calculations in this white paper we have assumed a conservative 32~GHz bandwidth, but larger bandwidths will likely be technically possible across most bands (for example, SCUBA-2 bandwidths are $\sim$40--50\,GHz), but limited by the atmospheric transmission windows available.

A large field of view is needed for the continuum camera to facilitate our science cases. It is anticipated that the first-generation continuum camera can have a 1~deg$^2$ field of view, significantly larger than current instruments (e.g., $\sim 13$ arcmin$^2$ for TolTEC or $\sim$8 arcmin$^2$ for SCUBA-2). With an angular resolution of $15''$ to $1.6''$ from Band 3 to Band 10 (Table~\ref{tab:atlast-angular-resolutions}), this means a camera would require at least a $2000 \times 2000$ pixels to sample the point spread function by a factor of $\gtrsim 1$--8. Alternatively, the mapping needs could be met with a sparser array with $\sim10^6$ pixels that undersamples the point spread function and a scanning or `jiggle' observing strategy.  A large field of view will also be crucial for observing emission on large spatial scales. Filtering of the variable sky-background emission will restrict the sensitivity to scales at most as large as the field of view (while not requiring fully sampled arrays).

\subsection{Submillimeter Integral Field Spectrograph}

A submillimeter IFU will be a remarkable game-changer for large-area surveys. A high spectral resolution of $R \gtrsim 1000$ is needed in order to separate lines. A field of view of 1~deg$^2$ with a spaxel spacing of about $4''$ will be critical for the mapping speed and Nyquist sampling of the beams. This means an IFU pixel array of $\sim 900 \times 900$, and a channel number of 1000. A bandwidth of at least 32~GHz is required for the aforementioned surveys. Equipping AtLAST with IFUs at each of Band~6 (198--230~GHz) and Band~3 (83--115~GHz) will be complementary and give maximum flexibility to accommodate varying weather conditions.

\subsection{Multi-beam Heterodyne Spectrometer}

For the nearby galaxy spectral line survey (Sec.~\ref{sec: line survey}), in the case where a high spectral resolution ($R \gtrsim 1000$) IFU is unavailable, a multi-beam heterodyne will be critical to achieve the goal of submillimeter line survey in nearby galaxies. As shown in Fig.~\ref{fig:heterodyne}, the number of beam units is important to reduce the required observing time to cover the whole galaxies. The spectral resolution requirement for most extragalactic targets is modest (compared to the needs for Galactic science), but the AtLAST heterodyne array should ideally have  $\lesssim 0.5$ km/s resolution, to enable additional kinematic studies in the LMC, SMC and other very nearby galaxies.

\begin{table*}
    \centering\small
    \caption{Common extragalactic molecular/atomic lines.}
    \begin{tabular}{c c c c}
    \hline
    \hline
     Band & species \& transition & rest-frame frequency & resolution \\
    \hline
     Band 3 
        %& DCO$^+$ $J=1\text{--}0$ & 72.039 & $20''$ \\
        %& SiO & 86.243 & $17''$ \\
        %& SiO & 86.847 & $17''$ \\
        %& C$_2$H $J=3/2\text{--}1/2$ & 88.632 & $17''$ \\
        & HCN $J=1\text{--}0$ & 88.632 & $17''$ \\
        & HCO$^{+}$ $J=1\text{--}0$ & 89.189 & $17''$ \\
        & HNC $J=1\text{--}0$ & 90.664 & $17''$ \\
        & N$_2$H$^+$ $J=1\text{--}0$ & 93.174 & $16''$ \\
        & CS $J=2\text{--}1$ & 97.981 & $15''$ \\
        & C$^{18}$O $J=1\text{--}0$ & 109.782 & $14''$ \\
        & $^{13}$CO $J=1\text{--}0$ & 110.201 & $14''$ \\
        & C$^{17}$O $J=1\text{--}0$ & 112.359 & $13''$ \\
        & CN $J=3/2\text{--}1/2$ & 113.491 & $13''$ \\
        & CO $J=1\text{--}0$ & 115.271 & $13''$ \\
    \hline
     Band 4 
        & H$_2$CO $2_{1,2}\text{--}1_{1,1}$	 & 140.840 & $11''$ \\
        & CS $J=3\text{--}2$ & 146.969 & $10''$ \\
    \hline
     Band 5 
        & HCN $J=2\text{--}1$ & 177.261 & $8''$ \\
        & HCO$^{+}$ $J=2\text{--}1$ & 178.375 & $8''$ \\
        & HNC $J=2\text{--}1$ & 181.325 & $8''$ \\
        & N$_2$H$^+$ $J=2\text{--}1$ & 186.345 & $8''$ \\
        & CS $J=4\text{--}3$ & 195.954 & $8''$ \\
    \hline
     Band 6 
        & C$^{18}$O $J=2\text{--}1$ & 219.560 & $7''$ \\
        & $^{13}$CO $J=2\text{--}1$ & 220.399 & $7''$ \\
        & C$^{17}$O $J=2\text{--}1$ & 224.714 & $7''$ \\
        & CO $J=2\text{--}1$ & 230.538 & $7''$ \\
        %& $^{13}$CO $J=2\text{--}1$ & 220.399 & $7''$ \\
        & CS $J=5\text{--}4$ & 244.935 & $6''$ \\
        & HCN $J=3\text{--}2$ & 265.886 & $6''$ \\
        & HCO$^{+}$ $J=3\text{--}2$ & 267.558 & $6''$ \\
        & HNC $J=3\text{--}2$ & 271.981 & $6''$ \\
        & N$_2$H$^+$ $J=3\text{--}2$ & 279.512 & $6''$ \\
    \hline
     Band 7 
        & CS $J=6\text{--}5$ & 293.912 & $5''$ \\
        & C$^{18}$O $J=3\text{--}2$ & 329.331 & $5''$ \\
        & $^{13}$CO $J=3\text{--}2$ & 330.588 & $4''$ \\
        & C$^{17}$O $J=3\text{--}2$ & 337.061 & $4''$ \\
        & CS $J=7\text{--}6$ & 342.883 & $4''$ \\
        & CO $J=3\text{--}2$ & 345.796 & $4''$ \\
        & HCN $J=4\text{--}3$ & 354.505 & $4''$ \\
        & HCO$^{+}$ $J=4\text{--}3$ & 356.734 & $4''$ \\
        & HNC $J=4\text{--}3$ & 362.630 & $4''$ \\
        & N$_2$H$^+$ $J=4\text{--}3$ & 372.673 & $4''$ \\
    \hline
     Band 8 
        & CO $J=4\text{--}3$ & 461.041 & $3''$ \\
        & [C\textsc{i}] ${}^{3}P_{1}\text{--}{}^{3}P_{0}$ & 492.161 & $3''$ \\
        & \multicolumn{3}{l}{(fainter lines omitted hereafter)} \\
    \hline
     Band 9 
        & CO $J=6\text{--}5$ & 691.473 & $2''$ \\
    \hline
     Band 10 
        & CO $J=7\text{--}6$ & 806.652 & $2''$ \\
        & [C\textsc{i}] ${}^{3}P_{2}\text{--}{}^{3}P_{1}$ & 809.342 & $2''$ \\
    \hline
    \hline
    \end{tabular}
    \caption*{%
    Notes: Line frequencies are based on \href{https://splatalogue.online/}{NRAO Splatalogue database}. Angular resolution is naively calculated as $1.22\,\lambda/D$. 
    \label{tab:spectral lines}
    }
\end{table*}

%%%%%%%%%%%%%%%%%%%%%%%%%%%%%%%%%%%%%%%%%%%%%%%%%%%%%%%%%%%%%%%%%
%%%%%%%%%%%%%%%%%%%%%%%%%%%%%%%%%%%%%%%%%%%%%%%%%%%%%%%%%%%%%%%%%
%%   CONCLUSIONS
%%%%%%%%%%%%%%%%%%%%%%%%%%%%%%%%%%%%%%%%%%%%%%%%%%%%%%%%%%%%%%%%%
%%%%%%%%%%%%%%%%%%%%%%%%%%%%%%%%%%%%%%%%%%%%%%%%%%%%%%%%%%%%%%%%%
\section{Summary and conclusions}
\label{sec: conclusion}

In this paper, we have explored the potential of AtLAST, a future 50-m single dish submm telescope, to significantly increase our understanding of the physics and chemistry of interstellar gas and dust, star formation and galaxy evolution through large-scale surveys of galaxies in the nearby Universe. This should be viewed in the context of the overall "scientific portfolio" of AtLAST; nearby galaxies allow us to extend to a broader range of environments the detailed analyses of the ISM in our own Galaxy (Klaassen et al. in prep.) while at the same time informing the study of galaxies in the distant universe (Lee et al. in prep., van Kampen et al. in prep.). Taken together, all these lines of investigations enabled by AtLAST will provide the solid observational framework required to make profound advances in our understanding of star formation and ISM physics across the Universe.   

The two main features of AtLAST that will enable breakthrough advances in our understanding of nearby galaxies are its the sensitivity to extended, low surface brightness emission, and its large field of view allowing for efficient surveying or large areas at unprecedented resolution and depth. For the purposes of this overview paper, we have focused on four nearby Universe science areas where AtLAST will be particularly impactful, and in each case discuss possible surveys and the instrumentation required to achieve transformational outcomes. 

Our first science case looks at two of the nearest neighbours of the Milky Way, the LMC and the SCM. An AtLAST continuum camera with a field of view of $\sim 1$ deg$^2$ could map the two galaxies fully in three bands (selected in the frequency range 100--680~GHz) with $\sim 2900$~h on-source integration; this observing time would be significantly reduced with a multi-band camera. The uniqueness of these observations would be the sensitivity to the diffuse, low surface brightness ISM, which is completely missed in current observations with both interferometers and 10-15m class single dish telescopes. This will allow us to answer crucial questions about the early stages of cloud formation, for example. 

We then identified the study of magnetic fields on galactic scales as another area where AtLAST can make a unique contribution. With AtLAST, we will be able to map the magnetic fields ($\vec{B}$) from dust polarization in possibly a hundred nearby galaxies, when currently only a handful of nearby galaxies have dust polarization observations (from the decommissioned SOFIA observatory and the 15-meter JCMT). Complementary to, but different from, the radio (L-/C-band) polarization, the submillimeter polarization uniquely probes the thermal emission of dust grains, and thus opens an extra dimension in understanding the complexity of the interplay between the $\vec{B}$ fields and star formation. AtLAST will achieve the largest sample of this kind at giant molecular cloud scales (tens to a hundred parsec) while also tracing the link between galactic- and cloud-scale magnetic fields. 

Our third and fourth science cases take advantage of the expected extremely large bandwidth and mapping speed of the AtLAST instruments to achieve the largest-ever samples for submillimeter spectral line and molecular gas surveys. Whereas the current largest submillimeter interferometer, ALMA, is capable of detecting hundreds of lines (e.g., the ALCHEMI survey) or observing the most important CO lines in slightly over a hundred nearby galaxies (e.g., the PHANGS survey), AtLAST will enable unprecedented statistical astrochemistry and CO luminosity/gas mass function studies by increasing the sample sizes by several orders of magnitude.  With such surveys, we will be able to answer fundamental questions regarding the impact of large scale environment on star formation efficiency and gas availability in galaxies, the shut-down of star formation during the quenching process, and the nature of the star formation efficiency variations across the local galaxy population, to name just a few.

\newpage

\begin{table*}
    \centering
    \caption{Angular scale (AR) and maximum recoverable scale (MRS)}
    \begin{tabular}{c | c | c |c c c}
    \hline
    \hline
        frequency & Band & AtLAST 50m & \multicolumn{3}{c}{ALMA 43$\times$12m AR/MRS} \\
              & & AR & C1 & C3 & C5 \\
        (GHz) & & (arcsec) & (arcsec) & (arcsec) & (arcsec) \\
    \hline
        100  & 3 (84--116) & 15.088 & 3.38$/$28.5 & 1.42$/$16.2 & 0.55$/$6.70 \\
        200  & 5 (163--211) & 7.544  & 1.69$/$14.3 & 0.71$/$8.10 & 0.28$/$3.35 \\
        300  & 7 (275--373) & 5.029  & 1.13$/$9.50 & 0.47$/$5.40 & 0.18$/$2.23 \\
        400  & 8 (385--500) & 3.772  & 0.84$/$7.12 & 0.35$/$4.05 & 0.14$/$1.68 \\
        500  & 8 (385--500) & 3.018  & 0.68$/$5.70 & 0.28$/$3.24 & 0.11$/$1.34 \\
        600  & 9 (602--720) & 2.515  & 0.56$/$4.75 & 0.24$/$2.70 & 0.092$/$1.12 \\
        700  & 9 (602--720) & 2.155  & 0.48$/$4.07 & 0.20$/$2.31 & 0.079$/$0.96 \\
        800  & 10 (787--950) & 1.886  & 0.42$/$3.56 & 0.18$/$2.02 & 0.069$/$0.84 \\
        900  & 10 (787--950) & 1.676  & 0.38$/$3.17 & 0.16$/$1.80 & 0.061$/$0.74 \\
    \hline
    \end{tabular}\\
    \caption*{%
    %\begin{minipage}[t]{0.8\linewidth}
    Notes: AR is naively calculated as $1.22\,\lambda/D$, which may slightly deviate from the AtLAST 50-m angular resolution in reality. 
    ALMA (ACA) field of view: $\lesssim 0.08$ ($\lesssim 0.24$)~arcmin$^2$ (single-beam receiver) vs. AtLAST field of view 1--2~deg$^2$ (large-format continuum camera or multi-beam receiver) at 300~GHz. 
    ALMA collecting area: $43 \times \pi \times 6^2 \sim 4863 \, \mathrm{m^2}$ vs. ACA collecting area: $12 \times \pi \times 3.5^2 \sim \, 462 \mathrm{m^2}$ vs. AtLAST collecting area: $\pi \times 25^2 \sim 1963 \, \mathrm{m^2}$. 
    ALMA available time in the most-compact C1--C2 configurations for faint, diffuse emission: $\sim 50$~days vs. AtLAST available time: $\sim 330$~days. 
    To recover the zeroth-$uv$ data, namely ``total power'', the ALMA total power antenna 12m dish is $17\times$ less sensitive than the AtLAST 50m dish when only considering the area difference. 
    %\end{minipage}
    \label{tab:atlast-angular-resolutions}
    }
\end{table*}

\newpage

\section*{Data and software availability} 
The calculations used to derive integration times for this paper were done using the AtLAST sensitivity calculator, a deliverable of Horizon 2020 research project `Towards AtLAST', and available from \href{https://github.com/ukatc/AtLAST_sensitivity_calculator}{this link}.

\section*{Grant information}
This project has received funding from the European Union’s Horizon 2020 research and innovation program under grant agreement No 951815 (AtLAST).
L.D.M. acknowledges support by the French government, through the UCA\textsuperscript{J.E.D.I.} Investments in the Future project managed by the National Research Agency (ANR) with the reference number ANR-15-IDEX-01.
M.L. acknowledges support from the European Union’s Horizon Europe research and innovation programme under the Marie Sk\l odowska-Curie grant agreement No 101107795.
M.C. gratefully acknowledges funding from the DFG through an Emmy Noether Research Group (grant number CH2137/1-1).

\begingroup
\small%
\bibliographystyle{apj_mod}
\setlength{\parskip}{0pt}
\setlength{\bibsep}{0pt}

\endgroup

\end{document}